\documentstyle[axodraw,12pt]{article}
\makeatletter  

\@addtoreset{equation}{section} \makeatother

\newcommand{\hs}{\hspace*{0.5cm}}
\newcommand{\vs}{\vspace*{0.5cm}}
\newcommand{\be}{\begin{equation}}
\newcommand{\ee}{\end{equation}}
\newcommand{\bea}{\begin{eqnarray}}
\newcommand{\eea}{\end{eqnarray}}
\newcommand{\ben}{\begin{enumerate}}
\newcommand{\een}{\end{enumerate}}
\newcommand{\nn}{\nonumber}
\newcommand{\crn}{\nonumber \\}

\newcommand{\la}{\lambda}

\newcommand{\fr}{\frac}
\newcommand{\bc}{\begin{center}}
\newcommand{\ec}{\end{center}}

\def\lappeq{\mathrel{\rlap{\raise.5ex\hbox{$<$}}
{\lower.5ex\hbox{$\sim$}}}}
\begin{document}

\bc {\Large Neutrino Masses in the Supersymmetric \\
 $\mbox{SU}(3)_{C}\otimes \mbox{SU}(3)_{L} \otimes \mbox{U}(1)_{X}$ Model\\
 with right-handed neutrinos}\\
\vspace*{1cm}

{\bf P. V. Dong$^a$, D. T. Huong$^a$, M. C. Rodriguez$^b$}
and {\bf H. N. Long$^a$}\\

\vspace*{0.5cm}

$^a$ {\it Institute of Physics, VAST, P. O. Box 429, Bo Ho, Hanoi
10000, Vietnam}\\

 $^b$ {\it Funda\c{c}\~{a}o Universidade Federal do Rio Grande,
 Departamento de F\'\i sica,\\
 Av. It\'alia, km 8, Campus Carreiros,
 96201-900 Rio Grande, RS,  Brazil}

\ec

\begin{abstract}
The R-symmetry formalism is applied for the supersymmetric
$\mbox{SU}(3)_{C}\otimes \mbox{SU}(3)_{L} \otimes \mbox{U}(1)_{X}$
(3-3-1) model with right-handed neutrinos. For this kind of
models, we study generalization of the MSSM relation among
R-parity, spin and matter-parity. Discrete symmetries for the
proton stable in this model are imposed, and we show that in such
a case it is able to give leptons masses at only the tree level
contributions required. A simple mechanism for the mass generation
of the neutrinos is explored. We show that at the low-energy
effective theory, neutrino spectrum contains three Dirac fermions,
one massless and two degenerate in mass. At the energy-level where
the mixing among them with neutralinos turned on, neutrinos obtain
Majorana masses and correct the low-energy effective result which
naturally gives rise to an inverted hierarchy mass pattern. This
mass spectrum can fit the current data with minor fine-tuning.
Consistent values for masses of the charged leptons are also
given. In this model, the MSSM neutralinos and charginos can be
explicitly identified in terms of the new constraints on masses
which is not as in a supersymmetric version of the minimal 3-3-1
model.

\vs


\end{abstract}

\section{Introduction}

Although the Standard Model (SM) gives very good results in
explaining the observed properties of the charged fermions, it is
unlikely to be the ultimate theory. It maintains the masslessness
of the neutrinos to all orders in perturbation theory, and even
after non-pertubative effects are included. The recent
groundbreaking discovery of nonzero neutrino masses and
oscillations \cite{superk} has put massive neutrinos as one of
evidences on physics beyond the SM.

The Super-Kamiokande experiments on the atmospheric neutrino
oscillations have indicated to the difference of the squared
masses and the mixing angle with fair accuracy
\cite{hanoiconf,fogli}
\begin{eqnarray}
\label{eqn::atm}
\Delta m^2_{\mathrm{atm}}  & = & 1.3 \div 3.0 \times 10^{-3}  {\rm eV ^2}, \\
\sin^2 2 \theta_{\mathrm{atm}}  & > & 0.9.
\end{eqnarray}
While, those from the combined fit of the solar and reactor
neutrino data point to
\begin{eqnarray}
\label{eqn::solar}
\Delta m^2_{\odot} ~ & = &
8.0^{+ 0.6}_{- 0.4} \times 10^{-5} ~ \rm{ eV^2}, \\
\tan ^2 \theta_{\odot} ~ & = & 0.45^{+0.09}_{-0.07}.
\end{eqnarray}
Since the data provide only the information about the differences
in $m_{\nu}^2$, the neutrino mass pattern can be either almost
degenerate or hierarchical. Among the hierarchical possibilities,
there are two types of normal and inverted hierarchies. In the
literature, most of the cases explore normal hierarchical one in
each. In this paper, we will mention on a supersymmetric model
which naturally gives rise to three pseudo-Dirac neutrinos with an
inverted hierarchical mass pattern.

The gauge symmetry of the SM as well as those of many extensional
models by themselves fix only the gauge bosons. The fermions and
Higgs contents have to be chosen somewhat arbitrarily. In the SM,
these choices are made in such a way that the neutrinos are
massless as mentioned. However, there are other choices based on
the SM symmetry that neutrinos become massive. We know these from
the popular seesaw \cite{seesaw} and radiative \cite{rad} models.
Particularly, the models based on the $\mbox{SU}(3)_C\otimes
\mbox{SU}(3)_L \otimes \mbox{U}(1)_X$ gauge unification group
\cite{ppf,flt,recent}, called 3-3-1 models, give more stricter
fermion contents. Indeed, only three fermion generations are
acquired as a result of the anomaly cancellation and the condition
of QCD asymptotic freedom. The arbitrariness in this case are only
behind which SM singlets put in the bottoms of the lepton
triplets? In some scenarios, exotic leptons may exist in the
singlets. Result of this is quite similar the case of the SM
neutrinos. As a fact, the mechanisms of the Zee's type \cite{rad}
for neutrino masses arise which been explored in Ref.\cite{yakip}.

Forbidding the exotic leptons, there are two main versions of the
3-3-1 models as far as minimal lepton sectors is concerned. In one of
them \cite{ppf} the three known left-handed lepton components for
each generation are associated to three $\mbox{SU}(3)_L$ triplets
as $(\nu_l,l,l^c)_L$, in which $l^c_L$ is related to the
right-handed isospin singlet of the charged lepton $l$ in the SM.
No extra leptons are needed and therefore it calls that a minimal
3-3-1 model. In the variant model \cite{flt} three
$\mbox{SU}(3)_L$ lepton triplets are of the form $(\nu_l, l,
\nu_l^c)_L$, where $\nu_l^c$ is related to the right-handed
component of the neutrino field $\nu_l$, thus called a model with
the right-handed neutrinos. This kind of the 3-3-1 models requires
only a more economical Higgs sector for breaking the gauge
symmetry and generating the fermion masses. Among the new gauge
bosons in this model, the neutral non-Hermitian bilepton field
$X^0$ may give promising signature in accelerator experiments and
may be also the source of neutrino oscillations~\cite{til}. In the
current paper, the neutrinos of the 3-3-1 model with right-handed
neutrinos is a subject for extended study.

The 3-3-1 model with right-handed neutrinos gives the tree level
neutrino mass spectrum with three Dirac fermions, one massless and
two degenerate in mass \cite{changlong}. This is clearly not
realistic under the experimental data. However, this pattern may
be severely changed by quantum effects and gives rise to an
inverted hierarchy mass pattern. This is a {\it specific} feature
of the 3-3-1 model with right-handed neutrinos which was
considered in Ref.\cite{changlong} (see also \cite{dias}), but
such effects exist in the very high level of the loop corrections.

Some years ago, one of us was proposed the construction of the
supersymmetric 3-3-1 model with right-handed neutrino
\cite{susy3312}. In this paper, we explore a consistent neutrino
mass spectrum of such a type but in a different side. Namely, that
tree level mass spectrum will become a real one of the massive
neutrinos due to mixing among them with the neutralinos in the
supersymmetric version of the model \cite{susy3312}. In this case,
we show that an inverted hierarchy mass spectrum for the neutrinos
may be obtained but at only required the tree level contribution
which can fit the current data with some minor fine-tuning. Thus,
our result differs from many extensions of the SM. As far as the
mechanism concerned, it obviously keeps in the kind of an seesaw
one. It is not as in the case of the minimal 3-3-1 model, in which
its supersymmetric version \cite{331susy} gives only the real
lepton mass spectra when the one loop corrections added
\cite{lepmass}. Moreover, in our case the charged leptons always
gain consistent masses from different impacts due to mixing among
the neutrinos with the neutralinos.

The outline of this work is as follows. In Sec.
\ref{sec:rsymmetry} we review the concept of R-symmetry and
R-parity, in order to apply this concept on the supersymmetric
3-3-1 model with right-handed neutrinos. In Sec.
\ref{sec:rparitysusy3312} we define the R-charge in our model in
order to get similar results as in the Minimal Supersymmetric
Standard Model (MSSM). While in Sec. \ref{sec:neutrinosmassterms}
we impose another discrete symmetry that allow neutrino masses but
forbid the proton decay. On Sec. \ref{sec:fermionmasses} we
calculate the fermion masses in our model. Our conclusions are
found in the last section. At least, in Appendix \ref{neumatel},
we present the mass matrix elements of the neutral fermions.

\section{R Symmetry}
\label{sec:rsymmetry}

It is important to note that the SM can explain the conservation
of lepton number ($L$) and of baryon number ($B$) without needing
to any discrete symmetry. However, this is not the case of
supersymmetric theories where only if interactions of conserving
both $L$ and $B$ are required, one has to impose one discrete
symmetry. This section recalls how R-parity emerged as a discrete
remnant of continuous $\mathrm{U}(1)$ R-symmetry which is
necessarily broken so that the gauginos and gluinos to acquire
masses in the MSSM.

\subsection{R Symmetry in Superspace Formalism}

The R-symmetry was introduced in 1975 by A. Salam and J. Strathdee
\cite{r1} and in an independent way by P. Fayet \cite{r2} to avoid
the interactions that violate either lepton number or baryon
number. There is very nice review about this subject in
Refs.\cite{barbier,moreau}.

The concept of R-symmetry is better understood in superspace
formalism, where the R-symmetry is a $\mathrm{U}(1)$ continuous
symmetry, parametrized by $\alpha$. The operator which produces
this symmetry is going to be denoted as ${\bf R}$. This operator
acts on the superspace coordinate $\theta$, $\bar{\theta}$ as
follows \cite{bailin} \bea  {\bf R} \theta &\rightarrow& e^{-i
\alpha} \theta, \crn {\bf R} \bar{\theta} &\rightarrow& e^{i
\alpha} \bar{\theta}. \label{The R-Invariance prop 1} \eea Hence
the $\theta$ has R-charge to be  $\mathrm{R}( \theta )= -1$, while
$\bar{\theta}$ is $\mathrm{R}( \bar{\theta} )=1$.

The operator ${\bf R}$ acts on chiral superfields
$\Phi(x,\theta,\bar{\theta})$ and anti-chiral superfields
$\bar{\Phi}(x,\theta,\bar{\theta})$, respectively, in the
following way~\cite{wb} \bea
   {\bf R} \Phi(x,\theta,\bar{\theta}) &=& e^{2 i n_{\Phi}\alpha}
\Phi(x, e^{-i\alpha}\theta ,e^{i\alpha}\bar{\theta} ),\label{eq3}\\
   {\bf R} \bar{ \Phi}(x,\theta,\bar{\theta}) &=&
          e^{-2 i n_{\Phi}\alpha}\bar{ \Phi}(x, e^{-i\alpha}\theta ,
          e^{i\alpha}\bar{\theta} ),
          \label{The R-Invariance prop 2}
\eea where $2n_{ \Phi}$ is the R-charge of the above chiral
superfield. This new charge $n_{\Phi}$ is an additive conserved
quantum number. This operator acts on the vectorial superfield  by
the rule \bea
   {\bf R} V(x,\theta,\bar{\theta}) &=&
V(x, e^{-i\alpha}\theta , e^{i\alpha}\bar{\theta} ).
         \label{The R-Invariance prop 3}
\eea

The expansion of the superfields in terms of  $\theta$ and
$\bar{\theta}$, see \cite{wb}, is given by \bea \Phi(x, \theta,
\bar{ \theta}) &=& A(x) + \sqrt{2} \theta \psi (x) + \theta \theta
F(x) \crn & & + i  \theta \sigma^m \bar{ \theta} \partial_m A(x) -
\frac{i}{ \sqrt{2}}( \theta \theta ) \partial_m \psi (x) \sigma^m
\bar{ \theta} \crn & &+ \frac{1}{4} (\theta \theta)( \bar{ \theta}
\bar{ \theta} ) \Box A(x),\label{eq1} \\ \bar{ \Phi}(x, \theta,
\bar{ \theta}) &=& \bar{A}(x) + \sqrt{2} \bar{ \theta} \bar{ \psi}
(x) + \bar{ \theta} \bar{ \theta} \bar{F}(x) \crn & &-i \theta
\sigma^m \bar{ \theta} \partial_m \bar{A}(x) + \frac{i}{
\sqrt{2}}( \bar{ \theta} \bar{ \theta} ) \theta \sigma^m
\partial_m \bar{ \psi} (x) \crn & &+ \frac{1}{4} (\theta \theta)(
\bar{ \theta} \bar{ \theta} ) \Box \bar{A}(x),\label{eq2}\\
V_{WZ}(x, \theta, \bar{ \theta}) &=& -\theta \sigma^{m} \bar{
\theta} A_{m}(x) + i( \theta \theta ) \bar{ \theta} \bar{
\lambda}(x) - i(\bar{ \theta} \bar{ \theta} )\theta \lambda(x)
\crn & &+ \frac{1}{2} (\theta \theta )( \bar{ \theta} \bar{
\theta} )D(x),\ \mbox{expansion in Wess-Zumino gauge},
\label{exp1}\eea
 where $A(x),F(x)$ and $D(x)$ are scalar fields;
$\psi(x)$ and $\lambda(x)$ are fermion fields, while $A_{m}(x)$ is
 vector field.

Combining  Eqs.(\ref{eq3}) and (\ref{eq1}) we get the following
transformations for the field components, respectively \bea \left.
\begin{array}{lcr}
A(x)    &\stackrel{{\bf R}}{\longmapsto}&       e^{2in_{\Phi}\alpha} A(x) \\
\psi (x) &\stackrel{{\bf R}}{\longmapsto}&e^{2i
\left( n_{\Phi}-\frac{1}{2} \right) \alpha}\psi (x) \\
F(x)    &\stackrel{{\bf R}}{\longmapsto}&       e^{2i \left( n_{\Phi}-1
\right)\alpha} F(x)
             \end{array} \right\}.
           \label{The R-Invariance prop 4a}
\eea Similarly, for the anti-chiral superfield we get\bea \left.
\begin{array}{lcr}
\bar{A}(x)   \stackrel{{\bf R}}{\longmapsto}
 e^{-2in_{\Phi}\alpha} \bar{A}(x) \\
\bar{\psi}(x) \stackrel{{\bf R}}{\longmapsto}
e^{-2i \left( n_{\Phi}-\frac{1}{2}
\right) \alpha}
\bar{\psi}(x) \\
\bar{F}(x)   \stackrel{{\bf R}}{\longmapsto}        e^{-2i \left( n_{\Phi}-1
\right)\alpha} \bar{F}(x)
             \end{array} \right\}.
           \label{The R-Invariance prop 4b}
\eea
 From Eqs.(\ref{The R-Invariance prop 3}) and (\ref{exp1}), the  field
components in the vector superfield transform as \bea
   \left.  \begin{array}{lcr}
A_{m}(x) &\stackrel{{\bf R}}{\longmapsto}&       A_{m}(x) \\
\lambda (x)     &\stackrel{{\bf R}}{\longmapsto}&
    e^{i\alpha} \lambda (x) \\
\bar{\lambda}(x)     &\stackrel{{\bf R}}{\longmapsto}&
    e^{-i\alpha} \bar{\lambda}(x) \\
D(x)       &\stackrel{{\bf R}}{\longmapsto}&       D(x)
          \end{array} \right\}.
          \label{The R-Invariance prop 5}
\eea The transformations in  Eqs.(\ref{The R-Invariance prop 4a}),
(\ref{The R-Invariance prop 4b}) and (\ref{The R-Invariance prop
5}) can be rewritten  in terms of 4-components spinors as
\cite{moreau} \be
\begin{array}{rcl}
A_{m}(x) &\stackrel{{\bf R}}{\longmapsto}  & A_{m}(x), \\
\Lambda(x) &\stackrel{{\bf R}}{\longmapsto}  & e^{i \gamma_5 \alpha}
\ \Lambda(x), \\
D(x) &\stackrel{{\bf R}}{\longmapsto}  & D(x),
\end{array}
\begin{array}{rcl}
A(x) &\stackrel{{\bf R}}{\longmapsto}  & e^{2in_{\Phi} \alpha} \ A(x), \\
\bar{A}(x) &\stackrel{{\bf R}}{\longmapsto}  &
e^{-2in_{\Phi} \alpha} \ \bar{A}(x), \\
\Psi(x) &\stackrel{{\bf R}}{\longmapsto}  & e^{2i \gamma_5
(n_{\Phi}-1/2)\alpha} \ \Psi(x), \\
F(x) &\stackrel{{\bf R}}{\longmapsto}  & e^{2i(n_{\Phi}-1) \alpha} \ F(x), \\
\bar{F}(x) &\stackrel{{\bf R}}{\longmapsto}  &
e^{-2i(n_{\Phi}-1) \alpha} \ \bar{F}(x).
\end{array}
\label{Rsym4} \ee
In Eq.(\ref{Rsym4}), $\Lambda$ that is the Majorana spinor
represents the gauginos, while $\Psi(x)$ represents the Dirac
spinor for quarks and leptons.

For products of left-handed chiral superfields, it is to be noted
that \bea
  {\bf R} \prod_{a}\;\Phi_{a}(x,\theta ,\bar{\theta})
    &=& e^{2i\sum_{a}n_{a}\alpha}\,\prod_{a}\Phi_{a}
    (x, e^{-i\alpha}\theta ,e^{i\alpha}\bar{\theta} ).
      \nn
\eea  Thus, the general superfield terms given below \bea
   & &   \int d^{4}\theta\; \bar{ \Phi}(x,\theta ,
   \bar{\theta}) \Phi(x,\theta ,\bar{\theta}), \crn
   & &   \int d^{4}\theta\; \bar{ \Phi}(x,\theta ,
   \bar{\theta}) e^{V(x,\theta ,\bar{\theta})}
   \Phi(x,\theta ,\bar{\theta}),\crn
   & &   \int d^{2}\theta\; \prod_{a}\Phi_{a}(x,\theta ,\bar{\theta}),
              \hspace{2cm} \mbox{if  } \sum_{a} n_{a} = 1,
              \label{invrparity}
\eea are all R-invariant.

\subsection{Continuous R-Symmetry in MSSM}

In the  MSSM~\cite{mssm} the left-handed fermions are in doublets,
whereas the right-handed antifermions are in singlets:
$\hat{L}_{L}\sim(1,{\bf2},-1)$, $\hat{l}^{c}_{L}\sim(1,{\bf1},2)$
and $\hat{Q}_{L}\sim(3,{\bf2},1/3)$,
$\hat{u}^{c}_{L}\sim(3^{*},{\bf1},-4/3)$,
$\hat{d}^{c}_{L}\sim(3^{*},{\bf1},2/3)$. The Higgs bosons are in
doublets, $\hat{H}_{1}\sim(1,{\bf2},-1)$ and
$\hat{H}_{2}\sim(1,{\bf2},1)$. With these multiplets, the
superpotential of the model is written as
\begin{eqnarray}
W_{2}&=&\mu\epsilon\hat{H}_{1}\hat{H}_{2}+ \mu_{0a}\epsilon
\hat{L}_{aL}\hat{H}_{2},\crn
W_{3}&=&f^{l}_{ab}\epsilon\hat{L}_{aL}\hat{H}_{1}\hat{l}^{c}_{bL}+
f^{u}_{ij}\epsilon\hat{Q}_{iL}\hat{H}_{2}\hat{u}^{c}_{jL} +
f^{d}_{ij}\epsilon\hat{Q}_{iL}\hat{H}_{1}\hat{d}^{c}_{jL}\crn & &
+ \lambda_{abc}\epsilon\hat{L}_{aL}\hat{L}_{bL}\hat{l}^{c}_{cL}+
\lambda^{\prime}_{iaj}\epsilon\hat{Q}_{iL}\hat{L}_{aL}\hat{d}^{c}_{jL}+
\lambda^{\prime\prime}_{ijk}\hat{d}^{c}_{iL}\hat{u}^{c}_{jL}\hat{d}^{c}_{kL}.
\label{mssmrpc}
\end{eqnarray}
Hereafter, the  superscript $L$ will be removed from the
superfields, and the $\mathrm{SU}(2)$ indices are default. The
superscript $^c$ indicates the charge conjugation and $\epsilon$
is the antisymmetric $\mathrm{SU}(2)$ tensor. The sub-indices $a,
b, c$ run over the lepton generations $e, \mu, \tau$ and  $i, j, k
=1, 2, 3$ run over the quark ones.

Because of Eqs.(\ref{eq3}) and (\ref{The R-Invariance prop 2}),
the terms proportional to $\lambda, \lambda^{\prime}$ and
$\lambda^{\prime \prime}$ are forbidden by the R-symmetry. The
following example illustrates this statement. Suppose that \bea
n_{H_{1}}&=&n_{H_{2}}=0,  \,\ \hbox{for} \,\ H_1,\ H_2, \crn
n_{Q}&=&n_{u}=n_{d}=n_{L}=n_{l}= \frac{1}{2}, \,\ \hbox{for} \,\
Q,\ u^c,\ d^c,\ L,\ l^c,\label{asign} \eea which imply \bea
\hat{H}_{1,2}(x,\theta , \bar{\theta}) &\stackrel{{\bf
R}}{\longmapsto}& \hat{H}_{1,2}(x,e^{-i \alpha}\theta ,e^{i
\alpha}\bar{\theta}), \label{Rfay1b} \eea
\bea \hat{\Phi}(x,\theta , \bar{\theta}) &\stackrel{{\bf
R}}{\longmapsto}& e^{i \alpha} \ \hat{\Phi}(x,e^{-i \alpha}\theta
, e^{i \alpha}\bar{\theta}),
 \ \hat{\Phi}=Q,\ u^{c},\ d^{c},\ L,\ l^{c}.
\label{Rfay1c} \eea By Eqs.(\ref{The R-Invariance prop 4a}) and
(\ref{The R-Invariance prop 4b}), their components transform as
\be
\begin{array}{rcl}
 H_{1,2}(x) &\stackrel{{\bf R}}{\longmapsto}& H_{1,2}(x), \\
 \tilde{H}_{1,2}(x) &\stackrel{{\bf R}}{\longmapsto}&
 e^{-i \alpha} \ \tilde{H}_{1,2}(x), \\
\tilde{f}_L(x) &\stackrel{{\bf R}}{
\longmapsto}& e^{i  \alpha} \ \tilde{f}_L(x), \\
 \tilde{f}^{c}_L(x) &\stackrel{{\bf R}}{
 \longmapsto}& e^{-i  \alpha} \ \tilde{f}^{c}_L(x), \\
 \Psi(x) &\stackrel{{\bf R}}{\longmapsto}&  \Psi(x).
\end{array}
\label{Rfay2}
\ee
We recall that $H_{1,2}(x)$ are the Higgs bosons,
$\tilde{H}_{1,2}(x)$ are the higgsinos, $\tilde{f}$ are the
squarks and sleptons, and $\Psi(x)$ are the quarks and leptons.
The consequence of the above transformation is summarized as \bea
\begin{array}{rcl}
\mbox{ordinary particle} &\stackrel{{\bf R}}{
\longmapsto}& \mbox{ordinary particle}, \\
\mbox{supersymmetric partner} &\stackrel{{\bf R}}{
\longmapsto}& e^{\pm i \alpha}
\mbox{supersymmetric partner}.
\end{array}
\eea

Under the transformation law in Eq.(\ref{Rfay2}), the conserving
terms  are given by \be W= f^{l}\hat{L}\hat{H_{1}}\hat{l}^{c}+
f^{d}\hat{Q}\hat{H_{1}}\hat{d}^{c}+
f^{u}\hat{Q}\hat{H_{2}}\hat{u}^{c}. \ee Therefore, the couplings
$\lambda, \lambda^{\prime}$ and $\lambda^{\prime \prime}$ are
forbidden by the charge assignment given in  Eq.(\ref{asign}).
These terms if they were allowed would induce the rapid proton
decay.  We allow only the terms from which the fermions in the model
gain masses~\cite{barbier}.

\subsection{Problem with Continuous R-Symmetry, Discrete R-Parity}

Because of Eq.(\ref{invrparity}), all the Lagrangians are
invariant under the continuous R-symmetry and this obviously
avoids the proton decay. However, such an unbroken continuous
R-symmetry which acts on the gaugino and gluino mass terms would
maintain them massless, even after a spontaneous breaking of the
supersymmetry. To see this, let us remember that the gaugino's
mass term is given by \cite{10} \be m_{\lambda} \left( \lambda
\lambda + \bar{\lambda} \bar{\lambda} \right), \label{gaugino mass
term} \ee which, under the R-symmetry (\ref{The R-Invariance prop
5}), transforms as \be m_{\lambda} \left( e^{2i \alpha}\lambda
\lambda + e^{-2i \alpha}\bar{\lambda} \bar{\lambda} \right). \ee
As a result, the mass term (\ref{gaugino mass term}) is not
invariant under the R-symmetry. This fact forces us to abandon the
continuous R-symmetry, in favour of the discrete R-symmetry,
called {\it R-parity}. Thereby the R-parity automatically allows
gluinos and other gauginos masses.

The discrete R-parity, denoted by ${\bf R}_{d}$, which is able to
solve the above problem can be obtained by putting $\alpha = \pi$.
Taking this value into account on Eqs.(\ref{The R-Invariance prop
1}), (\ref{eq3}), (\ref{The R-Invariance prop 2}) and (\ref{The
R-Invariance prop 3}) we get the following transformations \bea
{\bf R}_{d} \theta &\stackrel{{\bf R}_{d}}{\longmapsto}&- \theta ,
\crn {\bf R}_{d} \bar{\theta} &\stackrel{{\bf
R}_{d}}{\longmapsto}& - \bar{\theta},\crn  {\bf R}_{d}
\Phi(x,\theta,\bar{\theta}) &\stackrel{{\bf R}_{d}}{ \longmapsto}&
e^{2 i n_{\Phi}\pi}\Phi(x,- \theta ,- \bar{\theta} ),\crn {\bf
R}_{d} \bar{ \Phi}(x,\theta,\bar{\theta}) & \stackrel{{\bf
R}_{d}}{\longmapsto}& e^{-2 i n_{\Phi}\pi} \bar{ \Phi}(x,- \theta
,- \bar{\theta} ),\crn {\bf R}_{d} V(x,\theta,\bar{\theta})
&\stackrel{{\bf R}_{d}}{\longmapsto} & V(x,- \theta ,-
\bar{\theta} ). \label{invrparitydiscrete} \eea It is worth
emphasizing that, under this (discrete) transformation law, the
terms $\theta \theta$ and $\theta \theta \bar{\theta}
\bar{\theta}$ are invariants which is very helpful in further
analysis.

Now, under the discrete symmetry, the components of the
superfields transform as: \bea
   \left.  \begin{array}{lcr}
A(x) &\stackrel{{\bf R}_{d}}{\longmapsto}&       e^{2in_{\Phi}\pi} A(x) \\
\psi (x) &\stackrel{{\bf R}_{d}}{\longmapsto}& e^{2i
\left( n_{\Phi}-\frac{1}{2} \right) \pi}\psi (x)\\
F(x) &\stackrel{{\bf R}_{d}}{\longmapsto}& e^{2i \left( n_{\Phi}-1
\right) \pi} F(x)
             \end{array} \right\},
           \label{The R-discrete-parity}\\
   \left.  \begin{array}{lcr}
          A_{m}(x) &\stackrel{{\bf R}_{d}}{\longmapsto}&       A_{m}(x) \\
          \lambda (x)     &\stackrel{{\bf R}_{d}}{\longmapsto}&- \lambda (x) \\
          \bar{\lambda}(x)     &\stackrel{{\bf R}_{d}}{\longmapsto}&
          - \bar{\lambda}(x) \\
          D(x)       &\stackrel{{\bf R}_{d}}{\longmapsto}&       D (x)
          \end{array} \right\}.
          \label{The R-Invariance prop 5a}
\eea From (\ref{The R-Invariance prop 5a}), we see that
(\ref{gaugino mass term}) is, of course, invariant under the
discrete symmetry as mentioned. Moreover, the last term in
(\ref{invrparity}) can be redefined by \bea \int d^{2}\theta\;
\prod_{a}\Phi_{a}(x,\theta ,\bar{\theta}) \,\ , \hspace{2cm}
\mbox{if  } \sum_{a} n_{a} = 0. \label{invrparitydiscreta} \eea

In the next, we will show that there is a close connection between
R-parity and baryon, lepton number conservation laws. Its origin
is in our desire to get supersymmetric theories in which $B$ and
$L$ could be conserved, and simultaneously, to avoid unwanted
exchanges of spin-0 particles.

\subsection{Discrete R-Parity in MSSM}

Applying the conditions coming from (\ref{invrparitydiscreta}) on
Lagrangians in (\ref{mssmrpc}) we get the following equations \bea
n_{H_{1}}+n_{H_{2}}&=&0,\
n_{L}+n_{H_{2}}=0, \label{1a}\\
n_{H_{1}}+n_{L}+n_{l}&=&0,\
n_{H_{1}}+n_{Q}+n_{d}=0, \label{1b}\\
n_{H_{2}}+n_{Q}+n_{u}&=&0, \
n_{Q}+n_{L}+n_{d}=0, \label{1c}\\
2n_{L}+n_{l}&=&0, \,\ 2n_{d}+n_{u}=0 \label{1d}. \eea
 Unfortunately, not all of these relations can be satisfied
simultaneously. Only some of these constrains can be satisfied.
For example, choosing \bea n_{H_{1}}&=&0,\,\ n_{H_{2}}=0, \,\
n_{L}=\frac{1}{2}, \,\ n_{Q}=\frac{1}{2}, \crn n_{l}&=&-
\frac{1}{2}, n_{u}=-\frac{1}{2}, \,\ n_{d}=-\frac{1}{2},
\label{sym1} \eea the superfields will transform as \bea
\hat{V}(x,\theta,\bar \theta) &\stackrel{{\bf R}_{d}}{\longmapsto}
& \hat{V}(x,-\theta ,-\bar \theta ), \label{Rpa1a} \\
\hat{H}_{1,2}(x,\theta,\bar{\theta}) &\stackrel{{\bf
R}_{d}}{\longmapsto}& \hat{H}_{1,2}(x,-\theta,-\bar{\theta} ),
\label{Rpa1b} \\
 \hat{\Phi}(x,\theta,\bar{\theta}) &\stackrel{{\bf
R}_{d}}{\longmapsto}& - \hat{\Phi}(x,-\theta,-\bar{\theta} ),
 \ \Phi=Q,\ u^c,\ d^c,\ L,\ l^c.
\label{Rpa1c} \eea In terms of the field components, we obtain \be
\begin{array}{rcl}
A_{m}(x) &\stackrel{{\bf R}_{d}}{\longmapsto}& A_{m}(x), \\
\Lambda(x) &\stackrel{{\bf R}_{d}}{\longmapsto}& - \Lambda(x),
\end{array}
\begin{array}{rcl}
 H_{1,2}(x) &\stackrel{{\bf R}_{d}}{\longmapsto}& H_{1,2}(x), \\
 \tilde{H}_{1,2}(x) &\stackrel{{\bf R}_{d}}{\longmapsto}&
  - \tilde{H}_{1,2}(x),
\end{array}
\begin{array}{rcl}
 \tilde{f}(x) &\stackrel{{\bf R}_{d}}{\longmapsto}& - \tilde {f}(x), \\
 \Psi(x) &\stackrel{{\bf R}_{d}}{\longmapsto}&  \Psi(x).
\end{array}
\label{Rpa2} \ee
 Therefore,  the first condition in (\ref{1a}), both conditions in
(\ref{1b}), and again the first condition in (\ref{1c}) are
satisfied and this does not happen with the remaining conditions.
The terms in the superpotential (\ref{mssmrpc}) which satisfy the
rule with the parameters in (\ref{sym1}) are \be W=
\mu\epsilon\hat{H}_{1}\hat{H}_{2}+
f^{l}_{ab}\epsilon\hat{L}_{a}\hat{H}_{1}\hat{l}^{c}_{b}+
f^{u}_{ij}\epsilon\hat{Q}_{i}\hat{H}_{2}\hat{u}^{c}_{j} +
f^{d}_{ij}\epsilon\hat{Q}_{i}\hat{H}_{1}\hat{d}^{c}_{j}. \ee Their
others terms are forbidden which are behind that $\lambda$ and
$\lambda^{\prime}$ are kinds of the lepton number violating
parameters while $\lambda^{\prime \prime}$ is a type of the baryon
number violating parameter.

Equation (\ref{Rpa2}) suggests to classify the particles into two
types of so called R-even and R-odd. Here the R-even particles
$(R_d=+1)$ include the gluons, photon, $W^\pm$ and $Z$ gauge
bosons, the quarks, the leptons and the Higgs bosons. Whereas, the
R-odd particles $(R_d=-1)$ are their superpartners, i.e., the
gluinos, neutralinos, charginos, squarks and sleptons. Therefore,
R-parity is parity of R-charge of the continuous $\mathrm{U}(1)$
R-symmetry and defined by \be \label{eq:rp01}
\hbox{R-parity}=\left\{
\begin{array}{l}
+1\hs \hbox{for ordinary particles,} \vspace{2mm} \\
-1\hs \hbox{for their superpartners.}
\end{array}  \right.
\ee

The above intimate connection between R-parity and baryon number,
lepton number conservation laws can be made explicitly by
re-expressing (\ref{eq:rp01}) in terms of the spin $S$ and the
matter-parity $(-1)^{3B+L}$ as follows \cite{farrar78}:
\begin{equation}
\label{eq:rp02} \hbox{R-parity} = (-1)^{2S} (-1)^{3B+L}.
\end{equation}
Therefore, all scalar fields $(S=0)$ can be assigned $R$ values
\ben \item Usual scalars: $B=L=0 \Longrightarrow R=+1$ \item
Sleptons: $B=0,\ L=1 \Longrightarrow R=-1$ \item Squarks: $B=
\frac{1}{3},\ L=0 \Longrightarrow R=-1$\een Analogously for
fermions $(S=1/2)$ \ben \item Gauginos: $B=L=0 \Longrightarrow
R=-1$\item Leptons: $B=0,\ L=1 \Longrightarrow R=+1 $\item Quarks:
$B= \frac{1}{3},\ L=0 \Longrightarrow R=+1$\een Because of the
gauge bosons as well as all vectorial fields have $B=L=0$,
$\hbox{R-parity} =+1$. It is to be noted that the above assignment
is correct only for the MSSM, where the vector gauge bosons do not
carry the lepton number ($L=0$).

To finish this section, let us note that there will be a lot of
other choices of the charges in (\ref{1d}) to forbid the fast
proton decay \cite{Haki}. However, all such choices are due to the
action of R-symmetry which in a general way can be written as
\cite{rdiscreta}\bea \Phi \longrightarrow
e^{2in_{\Phi}\frac{2\pi}{N}}\Phi.\eea Here it is similar to a
$Z_N$ symmetry. Among those choices, there is a possibility which
allows neutrinos to gain masses. Indeed, choosing \bea
n_{H_{1}}&=&n_{H_{2}}=n_{L}=n_{l}=0, \crn n_{Q}&=& \frac{1}{2}, \
n_{u} = n_{d}=- \frac{1}{2}, \eea we get the following
transformation of the superfields\be (Q,u^c,d^c) \to -
(Q,u^c,d^c), \ (L,l^c,H_1,H_2) \to (L,l^c,H_1,H_2). \label{pabar}
\ee The terms which are allowed by this new R-parity are obtained
by \bea W&=& \mu\epsilon\hat{H}_{1}\hat{H}_{2}+
\mu_{0a}\hat{L}_{a}\hat{H}_{2}+
f^{l}_{ab}\epsilon\hat{L}_{a}\hat{H}_{1}\hat{l}^{c}_{b}+
f^{u}_{ij}\epsilon\hat{Q}_{i}\hat{H}_{2}\hat{u}^{c}_{j} +
f^{d}_{ij}\epsilon\hat{Q}_{i}\hat{H}_{1}\hat{d}^{c}_{j} \crn & & +
\lambda_{abc}\epsilon \hat{L}_{a}\hat{L}_{b}\hat{l}^{c}_{c}+
\lambda^{\prime}_{iaj}\epsilon
\hat{Q}_{i}\hat{L}_{a}\hat{d}^{c}_{j}. \eea
 As shown in Refs.~\cite{hall,banks,rv1}, this superpotential
 gives neutrinos masses.

\section{Discrete R-Parity in SUSY331RN}
\label{sec:rparitysusy3312}

In the supersymmetric 3-3-1 model with right-handed neutrinos
(SUSY331RN)~\cite{susy3312}, the fermionic  content is the
following: the left-handed fermions are in triplets/antitriplets
$L_{a}=(\nu_{a},l_{a}, \nu^{c}_{a})_L \sim({\bf1},{\bf3},-1/3)$,
$a=e,\ \mu,\ \tau$; $Q_{\alpha
L}=(d_{\alpha},u_{\alpha},d^{\prime}_{\alpha})
\sim({\bf3},{\bf3}^*,0)$, $\alpha=1,\ 2$,
$Q_{3L}=(u_{3},d_{3},u^{\prime})\sim({\bf3},{\bf3},1/3)$. The
right-handed components are in singlets:
$l^{c}_{a}\sim({\bf1},{\bf1},1)$, $u^{c}_{i},\ d^{c}_{i},\ i=1,\
2,\ 3$, which are similar to those in the SM. In addition,  the
exotic quarks  transform as  $u^{\prime c}\sim({\bf3}^*,
{\bf1},-2/3),\ d^{\prime c}_{\alpha}\sim({\bf3}^*,{\bf1},1/3)$.
The scalar content is minimally formed by three Higgs triplets:
$\eta=(\eta^{0}_{1},\eta^{-},\eta^{0}_{2})^T\sim({\bf1},{\bf3},-1/3)$;
$\chi=(\chi^{0}_{1},\chi^{-},\chi^{0}_{2})^T\sim({\bf1},{\bf3},-1/3)$
and
$\rho=(\rho^{+}_{1},\rho^{0},\rho^{+}_{2})^T\sim({\bf1},{\bf3},2/3)$.
The complete set of fields on the 331SUSYRN  is given in
Ref.~\cite{susy3312}.

In the  model under consideration, the superpotential is given by
\bea W_{2}&=&\mu_{0a}\hat{L}_{a} \hat{ \eta}^{\prime}+
\mu_{1a}\hat{L}_{a} \hat{ \chi}^{\prime}+ \mu_{ \eta} \hat{ \eta}
\hat{ \eta}^{\prime}+ \mu_{ \chi} \hat{ \chi} \hat{
\chi}^{\prime}+ \mu_{2} \hat{\eta} \hat{ \chi}^{\prime}+ \mu_{3}
\hat{\chi} \hat{ \eta}^{\prime}+ \mu_{ \rho} \hat{ \rho} \hat{
\rho}^{\prime}, \crn W_{3}&=& \lambda_{1ab} \hat{L}_{a} \hat{
\rho}^{\prime} \hat{l}^{c}_{b}+ \lambda_{2a} \epsilon \hat{L}_{a}
\hat{\chi} \hat{\rho}+ \lambda_{3a} \epsilon \hat{L}_{a}
\hat{\eta} \hat{\rho}+ \lambda_{4ab} \epsilon \hat{L}_{a}
\hat{L}_{b} \hat{\rho}+ \kappa_{1i} \hat{Q}_{3}
\hat{\eta}^{\prime} \hat{u}^{c}_{i}\crn & & + \kappa_{1}^{\prime}
\hat{Q}_{3} \hat{\eta}^{\prime} \hat{u}^{\prime c}+\kappa_{2i}
\hat{Q}_{3} \hat{\chi}^{\prime} \hat{u}^{c}_{i}+
\kappa_{2}^{\prime} \hat{Q}_{3} \hat{\chi}^{\prime}
\hat{u}^{\prime c}+ \kappa_{3\alpha i} \hat{Q}_{\alpha} \hat{\eta}
\hat{d}^{c}_{i}\crn & &+  \kappa_{3\alpha \beta}^{\prime}
\hat{Q}_{\alpha} \hat{\eta} \hat{d}^{\prime c}_{\beta}+
\kappa_{4\alpha i} \hat{Q}_{\alpha}\hat{\rho}\hat{u}^{c}_{i}+
\kappa_{4\alpha}^{\prime}
\hat{Q}_{\alpha}\hat{\rho}\hat{u}^{\prime c}
+\kappa_{5i}\hat{Q}_{3} \hat{\rho}^{\prime} \hat{d}^{c}_{i}+
\kappa_{5 \beta}^{\prime}\hat{Q}_{3} \hat{\rho}^{\prime}
\hat{d}^{c}_{\beta}\crn & &+ \kappa_{6\alpha i} \hat{Q}_{\alpha}
\hat{\chi} \hat{d}^{c}_{i}+ \kappa_{6\alpha \beta}^{\prime}
\hat{Q}_{\alpha} \hat{\chi} \hat{d}^{\prime c}_{\beta}+ f_{1}
\epsilon \hat{ \rho} \hat{ \chi} \hat{ \eta}+ f^{\prime}_{1}
\epsilon \hat{ \rho}^{\prime}\hat{ \chi}^{\prime}\hat{
\eta}^{\prime}+ \zeta_{\alpha \beta \gamma} \epsilon
\hat{Q}_{\alpha} \hat{Q}_{\beta} \hat{Q}_{\gamma}\crn & & +
\lambda^{\prime}_{\alpha ai}\hat{Q}_{\alpha}\hat{L}_{a}
\hat{d}^{c}_{i}+ \lambda^{\prime \prime}_{ijk} \hat{d}^{c}_{i}
\hat{u}^{c}_{j} \hat{d}^{c}_{k}+ \xi_{1ij \beta} \hat{d}^{c}_{i}
\hat{u}^{c}_{j} \hat{d}^{\prime c}_{\beta}+ \xi_{2 \alpha a
\beta}\hat{Q}_{\alpha} \hat{L}_{a} \hat{d}^{\prime c}_{\beta} \crn
& &+ \xi_{3i \beta} \hat{d}^{c}_{i} \hat{u}^{\prime c}
\hat{d}^{\prime c}_{\beta}+ \xi_{4ij} \hat{d}^{c}_{i}
\hat{u}^{\prime c} \hat{d}^{c}_{j}+ \xi_{5 \alpha i \beta}
\hat{d}^{\prime c}_{\alpha} \hat{u}^{c}_{i} \hat{d}^{\prime
c}_{\beta}+ \xi_{6 \alpha \beta} \hat{d}^{\prime c}_{\alpha}
\hat{u}^{\prime c} \hat{d}^{\prime c}_{\beta}. \label{sp3susy2}
\eea Applying the conditions coming from
(\ref{invrparitydiscreta}) on (\ref{sp3susy2}), we get the
following equations \bea n_{L}+n_{\eta^{\prime}}&=&0, \, \
n_{L}+n_{\chi^{\prime}}=0, \,\ n_{\eta}+n_{\eta^{\prime}}=0, \,\
n_{\chi}+n_{\chi^{\prime}}=0, \crn n_{\eta}+n_{\chi^{\prime}}&=&0,
\,\ n_{\chi}+n_{\eta^{\prime}}=0, \,\
n_{\rho}+n_{\rho^{\prime}}=0, \crn
n_{L}+n_{\rho^{\prime}}+n_{l}&=&0, \,\ n_{L}+n_{\chi}+n_{\rho}=0,
\,\ n_{L}+n_{\eta}+n_{\rho}=0, \,\ 2n_{L}+n_{\rho}=0, \crn
 n_{Q_{3}}+n_{\eta^{\prime}}+n_{u}&=&0, \,\
n_{Q_{3}}+n_{\eta^{\prime}}+n_{u^{\prime}}=0, \,\
n_{Q_{3}}+n_{\chi^{\prime}}+n_{u}=0, \crn
n_{Q_{3}}+n_{\chi^{\prime}}+n_{u^{\prime}}&=&0, \,\
n_{Q_{3}}+n_{\rho^{\prime}}+n_{d}=0, \,\
n_{Q_{3}}+n_{\rho^{\prime}}+n_{d^{\prime}}=0, \crn
n_{Q_{\alpha}}+n_{\eta}+n_{d}&=&0, \,\
n_{Q_{\alpha}}+n_{\eta}+n_{d^{\prime}}=0, \,\
n_{Q_{\alpha}}+n_{\chi}+n_{d}=0, \crn
n_{Q_{\alpha}}+n_{\chi}+n_{d^{\prime}}&=&0, \,\
n_{Q_{\alpha}}+n_{\rho}+n_{u}=0, \,\
n_{Q_{\alpha}}+n_{\rho}+n_{u^{\prime}}=0, \crn
3n_{Q_{\alpha}}&=&0, \,\ n_{\rho}+n_{\chi}+n_{\eta}=0, \,\
n_{\rho^{\prime}}+n_{\chi^{\prime}}+n_{\eta^{\prime}}=0, \crn
n_{Q_{\alpha}}+n_{L}+n_{d}&=&0, \,\
n_{Q_{\alpha}}+n_{L}+n_{d^{\prime}}=0,\crn 2n_{d}+n_{u}&=&0, \,\
n_{d}+n_{d^{\prime}}+n_{u}=0, \crn
n_{d}+n_{d^{\prime}}+n_{u^{\prime}}&=&0, \,\
2n_{d}+n_{u^{\prime}}=0, \,\ 2n_{d^{\prime}}+n_{u}=0, \crn
2n_{d^{\prime}}+n_{u^{\prime}}&=&0. \label{invchoosesusy2} \eea

Choosing the following R-charges \bea
n_{\eta}&=&n_{\eta^{\prime}}=n_{\chi}=n_{\chi^{\prime}}=n_{\rho}=
n_{\rho^{\prime}}=0, \crn n_{L}&=&n_{Q_{\alpha}}=n_{Q_{3}}=1/2,
\crn n_{l}&=&n_{u}=n_{d}=n_{u^{\prime}}=n_{d^{\prime}}=-1/2,
\label{rdiscsusy331rn} \eea and looking at Eq.(\ref{The
R-discrete-parity}), it is easy to see that all the fields $\eta$,
$\eta^{\prime}$, $\chi$, $\chi^{\prime}$, $\rho$, $\rho^{\prime}$,
$L$, $Q_{\alpha}$, $Q_{3}$, $l$, $u$, $u^{\prime}$, $d$ and
$d^{\prime}$ have R-charge equal to one, while their superpartners
have opposite R-charge similar to that in the MSSM. The terms
which satisfy the defined above symmetry (\ref{rdiscsusy331rn})
are \bea W&=& \frac{\mu_{ \eta}}{2} \hat{ \eta} \hat{
\eta}^{\prime}+ \frac{\mu_{ \chi}}{2} \hat{ \chi} \hat{
\chi}^{\prime}+ \frac{\mu_{ \rho}}{2} \hat{ \rho} \hat{
\rho}^{\prime}+ \frac{\mu_{2}}{2} \hat{\eta} \hat{ \chi}^{\prime}+
\frac{\mu_{3}}{2} \hat{\chi} \hat{ \eta}^{\prime}+ \frac{1}{3}
\left[ \lambda_{1ab} \hat{L}_{a} \hat{ \rho}^{\prime}
\hat{l}^{c}_{b} \right. \crn & &+  \kappa_{1i} \hat{Q}_{3}
\hat{\eta}^{\prime} \hat{u}^{c}_{i}+ \kappa_{1}^{\prime}
\hat{Q}_{3} \hat{\eta}^{\prime} \hat{u}^{\prime c}+ \kappa_{2i}
\hat{Q}_{3} \hat{\chi}^{\prime} \hat{u}^{c}_{i}+
\kappa_{2}^{\prime} \hat{Q}_{3} \hat{\chi}^{\prime}
\hat{u}^{\prime c}+ \kappa_{3\alpha i} \hat{Q}_{\alpha} \hat{\eta}
\hat{d}^{c}_{i} \crn & &+ \kappa_{3\alpha \beta}^{\prime}
\hat{Q}_{\alpha} \hat{\eta} \hat{d}^{\prime c}_{\beta}  +
\kappa_{4\alpha i} \hat{Q}_{\alpha}\hat{\rho}\hat{u}^{c}_{i}+
\kappa_{4\alpha}^{\prime} \hat{Q}_{\alpha}\hat{\rho}
\hat{u}^{\prime c}+ \kappa_{5i}\hat{Q}_{3} \hat{\rho}^{\prime}
\hat{d}^{c}_{i}+ \kappa_{5 \beta}^{\prime}\hat{Q}_{3}
\hat{\rho}^{\prime} \hat{d}^{c}_{\beta}\crn & &+  \kappa_{6\alpha
i} \hat{Q}_{\alpha} \hat{\chi} \hat{d}^{c}_{i}+ \kappa_{6\alpha
\beta}^{\prime} \hat{Q}_{\alpha} \hat{\chi} \hat{d}^{\prime
c}_{\beta}  \crn & & +\left. f_{1}\epsilon
\hat{\rho}\hat{\chi}\hat{\eta}+ f^{\prime}_{1}\epsilon
\hat{\rho}^{\prime}\hat{\chi}^{\prime}\hat{\eta}^{\prime} \right].
\label{rpartsusy331rn} \eea

Because of the lepton content of the considering model, the lepton
number $L$ obviously does not commute with the gauge symmetry.
However, a new conserved charge $\cal L$ can be constructed
through $L$ by making the linear combination $L= x\la_3 + y\la_8 +
{\cal L} I$ where $\la_3$ and $\la_8$ are the diagonal generators
of the $\mbox{SU}(3)_L$ group. Applying this operator on a lepton
triplet, the coefficients will be defined\be L =
\fr{2}{\sqrt{3}}\la_8 + {\cal L} I. \label{lepn} \ee Moreover, it
is useful to produce another conserved charge $\cal B$ which
itself is usual baryon number, $B ={\cal B} I$. Thus, the R-parity
in this model can be re-expressed via the spin $S$, new charges
$\mathcal{L}$ and $\mathcal{B}$ in terms of
\begin{equation}
\hbox{R-parity}=(-1)^{2S}(-1)^{3({\cal B}+{\cal L})},
\end{equation}
where the charges $ {\cal B}$ and ${\cal L}$ for the multiplets
are defined as follows \cite{changlong}
\begin{equation}
\begin{array}{|c|c|c|c|c|c|}
\hline
  Triplet & L & Q_{3} & \chi & \eta & \rho \\
  \hline
  {\cal B} \,\  charge & 0 & \frac{1}{3} & 0 & 0 & 0 \\ \hline
  {\cal L} \,\  charge & \frac{1}{3} &
  - \frac{2}{3} & \frac{4}{3} & - \frac{2}{3}
  & - \frac{2}{3} \\ \hline
\end{array}
\end{equation}
\begin{equation}
\begin{array}{|c|c|c|c|c|}
\hline
  Anti-Triplet & Q_{\alpha} & \chi^{\prime} & \eta^{\prime} & \rho^{\prime} \\
  \hline
  {\cal B} \,\  charge & \frac{1}{3} & 0 & 0 & 0 \\ \hline
  {\cal L} \,\  charge & \frac{2}{3} & - \frac{4}{3} & \frac{2}{3}
  & \frac{2}{3} \\ \hline
\end{array}
\end{equation}
\begin{equation}
\begin{array}{|c|c|c|c|c|c|}
\hline
  Singlet & l^{c} & u^{c} & d^{c} & u^{\prime c} & d^{\prime c} \\
  \hline
  {\cal B} \,\ charge & 0 & -\frac{1}{3} & -\frac{1}{3} &
  -\frac{1}{3} & -\frac{1}{3} \\ \hline
  {\cal L} \,\ charge & -1 & 0 &  0 & 2 & -2 \\ \hline
\end{array}
\end{equation}

From the superpotential given in Eq.(\ref{rpartsusy331rn}), it is
easy to see that the charged leptons gain mass only from the  term
\be -\frac{\lambda_{1ab}}{3}L_{a}
\rho^{\prime}l^{c}_{b}+hc.\label{shklcl} \ee The Higgs fields can
have VEVs given by \bea \langle\rho\rangle &=&(0,u,0)^T,\mbox{
}\langle\rho'\rangle=(0,u',0)^T,\crn \langle\eta\rangle
&=&(v,0,0)^T,\mbox{ }\langle\eta'\rangle=(v',0,0)^T, \crn
\langle\chi\rangle &=&(0,0,w)^T,\mbox{
}\langle\chi'\rangle=(0,0,w')^T.\label{vevsusy331rn}\eea Combining
(\ref{shklcl}) and (\ref{vevsusy331rn}) we get mass terms \be
-\frac{\lambda_{1ab}}{3}(l_{a}l^{c}_{b}+
\bar{l}_{a}\bar{l}^{c}_{b})u^{\prime},\ee which lead to the
following mass matrix  \bea X= \left(
\begin{array}{ccc}
  \fr{\la_{111}}{3}u' & \fr{\la_{112}}{3}u' & \fr{\la_{113}}{3}u' \\
  \fr{\la_{121}}{3}u' & \fr{\la_{122}}{3}u' & \fr{\la_{123}}{3}u' \\
  \fr{\la_{131}}{3}u' & \fr{\la_{132}}{3}u' & \fr{\la_{133}}{3}u'
\end{array}
\right).\nn\eea Hence, all the charged leptons get mass. Notice
that  only VEV of $\rho^{\prime}$ is enough to give  the charged
leptons  masses.

Remind that due to conservation of the $R$-parity defined in
Eq.(\ref{rdiscsusy331rn}), there are no terms which give neutrinos
masses in the superpotential. Thus, in this case the neutrinos
remain massless. From  another hand, the gaugino mass terms are
given by \be
 {\cal
L}_{GMT}=- \frac{m_{\lambda}}{2} \left[ \sum_{a=1}^{8}
(\lambda^{a}_{A}\lambda^{a}_{A}+\bar{\lambda}^{a}_{A}\bar{\lambda}^{a}_{A})
\right] -
\frac{m^{\prime}}{2}(\lambda_{B}\lambda_{B}+\bar{\lambda}_{B}\bar{\lambda}_{B}).
\label{gauginosmassterm} \ee Remind that in this  model the
non-Hermitian gauge bosons are defined as \bea \sqrt{2}\
W^{\pm}_{m}&=& V^{1}_{m} \mp iV^{2}_{m}, \crn
 \sqrt{2}\ Y^{\pm}_{m}&=& V^{6}_{m} \pm iV^{7}_{m}, \crn
 \sqrt{2}\ X^{0}_{m}&=& V^4_{m}- iV^{5}_{m}.
\eea According to these equations,  gauginos of the model are
defined as \bea \sqrt{2}\ \lambda^{\pm}_{W}&=& \lambda^{1}_{A} \mp
i \lambda^{2}_{A}, \crn \sqrt{2}\ \lambda^{\pm}_{Y}&=&
\lambda^{6}_{A} \pm i \lambda^{7}_{A}, \crn
 \sqrt{2}\ \lambda_{X^{0}}&=& \lambda^4_{A}-
i \lambda^{5}_{A}. \label{nonhermitiangauginos} \eea
 Then, in terms of these fields,  Eq.(\ref{gauginosmassterm})
 can be rewritten  as
\be -m_{\lambda}( \lambda^{-}_{W}\lambda^{+}_{W}+
\lambda^{-}_{Y}\lambda^{+}_{Y}+ \lambda_{X^{0}}\lambda_{X^{0*}})-
\frac{m_{\lambda}}{2}( \lambda^{3}_{A}\lambda^{3}_{A}+
\lambda^{8}_{A}\lambda^{8}_{A})- \frac{m^{\prime}}{2}
\lambda_{B}\lambda_{B}+hc \ee

The mass matrix of charginos and higgsinos arises from the
following Lagrangian \bea {\cal L}_{H \tilde{H}
\tilde{V}}&=&-\frac{ig}{\sqrt{2}} [
\bar{\tilde{\eta}}\lambda^{a}\eta \bar{\lambda}^{a}_{A}-
\bar{\eta}\lambda^{a}\tilde{\eta}\lambda^{a}_{A}+
\bar{\tilde{\rho}}\lambda^{a}\rho \bar{\lambda}^{a}_{A}-
\bar{\rho}\lambda^{a}\tilde{\rho}\lambda^{a}_{A}+
\bar{\tilde{\chi}}\lambda^{a}\chi \bar{\lambda}^{a}_{A}-
\bar{\chi}\lambda^{a}\tilde{\chi}\lambda^{a}_{A} \crn & &-
\bar{\tilde{\eta}}^{\prime}\lambda^{*a}\eta^{\prime}
\bar{\lambda}^{a}_{A}+
\bar{\eta}^{\prime}\lambda^{*a}\tilde{\eta}^{\prime}\lambda^{a}_{A}-
\bar{\tilde{\rho}}^{\prime}\lambda^{*a}\rho^{\prime}
\bar{\lambda}^{a}_{A}+
\bar{\rho}^{\prime}\lambda^{*a}\tilde{\rho}^{\prime}\lambda^{a}_{A}-
\bar{\tilde{\chi}}^{\prime}\lambda^{*a}\chi^{\prime}
\bar{\lambda}^{a}_{A}+
\bar{\chi}^{\prime}\lambda^{*a}\tilde{\chi}^{\prime}\lambda^{a}_{A}
]\crn  & &- \frac{ig^{\prime}}{\sqrt{2}} \left[ -
\frac{1}{3}(\bar{\tilde{\eta}}\eta
\bar{\lambda}_{B}-\bar{\eta}\tilde{\eta}\lambda_{B})-
\frac{1}{3}(\bar{\tilde{\chi}}\chi
\bar{\lambda}_{B}-\bar{\chi}\tilde{\chi}\lambda_{B})+
\frac{1}{3}(\bar{\tilde{\eta}}^{\prime}\eta^{\prime}
\bar{\lambda}_{B}-\bar{\eta}^{\prime}\tilde{\eta}^{\prime}\lambda_{B})
\right. \crn & & + \left.
\frac{1}{3}(\bar{\tilde{\chi}}^{\prime}\chi^{\prime}
\bar{\lambda}_{B}-\bar{\chi}^{\prime}\tilde{\chi}^{\prime}\lambda_{B})+
\frac{2}{3}(\bar{\tilde{\rho}}\rho
\bar{\lambda}_{B}-\bar{\rho}\tilde{\rho}\lambda_{B})-
\frac{2}{3}(\bar{\tilde{\rho}}^{\prime}\rho^{\prime}
\bar{\lambda}_{B}-\bar{\rho}^{\prime}\tilde{\rho}^{\prime}\lambda_{B})
\right]. \label{gauginosmassterm2}
\end{eqnarray}
Using Eq.(\ref{nonhermitiangauginos}), we can rewrite the
Lagrangian as \bea {\cal L}_{H \tilde{H} \tilde{V}}&=&+ig[v
\tilde{\eta}^{-}\lambda^{+}_{W}+w
\tilde{\chi}^{-}\lambda^{+}_{Y}+u(
\tilde{\rho}^{+}_{1}\lambda^{-}_{W}+\tilde{\rho}^{+}_{2}\lambda^{-}_{Y})-
v^{\prime} \tilde{\eta}^{\prime +}\lambda^{-}_{W}-w^{\prime}
\tilde{\chi}^{\prime +}\lambda^{-}_{Y} \crn &-& u^{\prime}(
\tilde{\rho}^{\prime -}_{1}\lambda^{+}_{W}+\tilde{\rho}^{\prime
-}_{2}\lambda^{+}_{Y})]- \frac{igv}{\sqrt{2}}\tilde{\eta}^{0}_{1}
\lambda^{3}_{A}-igv \tilde{\eta}^{0}_{2}\lambda_{X^{0}}-
\frac{igv}{\sqrt{2}}\tilde{\eta}^{0}_{1} \lambda^{8}_{A} \crn
&-&igw \tilde{\chi}^{0}_{1}\lambda_{X^{0*}}+ igw
\sqrt{\frac{2}{3}}\tilde{\chi}^{0}_{2} \lambda^{8}_{A}+
\frac{igu}{\sqrt{2}}\tilde{\rho}^{0} \lambda^{3}_{A}-
\frac{igu}{\sqrt{6}}\tilde{\rho}^{0} \lambda^{8}_{A}+
\frac{igv^{\prime}}{\sqrt{2}}\tilde{\eta}^{\prime 0}_{1}
\lambda^{3}_{A}+igv^{\prime} \tilde{\eta}^{\prime
0}_{2}\lambda_{X^{0*}} \crn &+&
\frac{igv^{\prime}}{\sqrt{6}}\tilde{\eta}^{\prime 0}_{1}
\lambda^{8}_{A}+igw^{\prime} \tilde{\chi}^{\prime
0}_{1}\lambda_{X^{0}}-igw^{\prime}
\sqrt{\frac{2}{3}}\tilde{\chi}^{\prime 0}_{2} \lambda^{8}_{A}+
\frac{igu^{\prime}}{\sqrt{6}}\tilde{\rho}^{\prime 0}
\lambda^{8}_{A}- \frac{igu^{\prime}}{\sqrt{2}}\tilde{\rho}^{\prime
0} \lambda^{3}_{A}-ig^{\prime}u \sqrt{\frac{2}{3}}
\tilde{\rho}^{0}\lambda_{B} \crn &+& \frac{ig^{\prime}v}{3
\sqrt{2}}\tilde{\eta}^{0}_{1}\lambda_{B} + \frac{ig^{\prime}w}{3
\sqrt{2}}\tilde{\chi}^{0}_{2}\lambda_{B}-\frac{ig^{\prime}v^{\prime}}{3
\sqrt{2}}\tilde{\eta}^{\prime 0}_{1}\lambda_{B} -
\frac{ig^{\prime}w^{\prime}}{3 \sqrt{2}}\tilde{\chi}^{\prime
0}_{2}\lambda_{B} +ig^{\prime}u^{\prime} \sqrt{\frac{2}{3}}
\tilde{\rho}^{0}\lambda_{B}+hc \eea
 The mass terms of the
higgsinos are given by \bea &-&
\frac{\mu_{\eta}}{2}\tilde{\eta}\tilde{\eta}^{\prime}-
\frac{\mu_{\chi}}{2}\tilde{\chi}\tilde{\chi}^{\prime}-
\frac{\mu_{\rho}}{2}\tilde{\rho}\tilde{\rho}^{\prime}-
\frac{f_{1}}{3}\epsilon(\tilde{\rho}\tilde{\chi}\eta + \rho
\tilde{\chi}\tilde{\eta}+ \tilde{\rho}\chi \tilde{\eta})-
\frac{f^{\prime}_{1}}{3}\epsilon(\tilde{\rho}^{\prime}
\tilde{\chi}^{\prime}\eta^{\prime} + \rho^{\prime}
\tilde{\chi}^{\prime}\tilde{\eta}^{\prime}+
\tilde{\rho}^{\prime}\chi^{\prime} \tilde{\eta}^{\prime}) +hc \nn
\end{eqnarray}
In the terms of field components, the above expression can be
rewritten as \bea &-&
\frac{\mu_{\eta}}{2}(\tilde{\eta}^{0}_{1}\tilde{\eta}^{\prime
0}_{1}+\tilde{\eta}^{-}\tilde{\eta}^{\prime +}
+\tilde{\eta}^{0}_{2}\tilde{\eta}^{\prime 0}_{2})-
\frac{\mu_{\rho}}{2}(\tilde{\rho}^{+}_{1}\tilde{\rho}^{\prime
-}_{1}+\tilde{\rho}^{0}\tilde{\rho}^{\prime 0}
+\tilde{\rho}^{+}_{2}\tilde{\rho}^{\prime -}_{2}) \crn &-&
\frac{\mu_{\chi}}{2}(\tilde{\chi}^{0}_{1}\tilde{\chi}^{\prime
0}_{1}+\tilde{\chi}^{-}\tilde{\chi}^{\prime +}
+\tilde{\chi}^{0}_{2}\tilde{\chi}^{\prime 0}_{2}) -
\frac{f_{1}}{3} \left[ ( \tilde{\rho}^{0}\tilde{\chi}^{0}_{2}-
\tilde{\rho}^{+}_{2}\tilde{\chi}^{-}) \eta^{0}_{1}+
(\tilde{\rho}^{+}_{1}\tilde{\chi}^{-}-
\tilde{\rho}^{0}\tilde{\chi}^{0}_{1}) \eta^{0}_{2} \right. \crn
&+& \left. ( \tilde{\chi}^{0}_{2}\tilde{\eta}^{0}_{1}-
\tilde{\chi}^{0}_{1}\tilde{\eta}^{0}_{2}) \rho^{0}+
(\tilde{\rho}^{+}_{2}\tilde{\eta}^{-}-
\tilde{\rho}^{0}\tilde{\eta}^{0}_{2}) \chi^{0}_{1}+
(\tilde{\rho}^{0}\tilde{\eta}^{0}_{1}-
\tilde{\rho}^{+}_{1}\tilde{\eta}^{-}) \chi^{0}_{2} \right]-
\frac{f^{\prime}_{1}}{3} \left[ (\tilde{\rho}^{\prime
0}\tilde{\chi}^{\prime 0}_{2}- \tilde{\rho}^{\prime
-}_{2}\tilde{\chi}^{\prime +})\eta^{\prime 0}_{1} \right. \crn &+&
\left. (\tilde{\rho}^{\prime -}_{1}\tilde{\chi}^{\prime +}-
\tilde{\rho}^{\prime 0}\tilde{\chi}^{\prime 0}_{1}) \eta^{\prime
0}_{2}+ (\tilde{\chi}^{\prime 0}_{2}\tilde{\eta}^{\prime 0}_{1}-
\tilde{\chi}^{\prime 0}_{1}\tilde{\eta}^{\prime 0}_{2})
\rho^{\prime 0}+ (\tilde{\rho}^{\prime -}_{2}\tilde{\eta}^{\prime
+}- \tilde{\rho}^{\prime 0}\tilde{\eta}^{ \prime 0}_{2})
\chi^{\prime 0}_{1} \right. \crn &+& \left. (\tilde{\rho}^{\prime
0}\tilde{\eta}^{\prime 0}_{1}- \tilde{\rho}^{\prime
-}_{1}\tilde{\eta}^{\prime +}) \chi^{\prime 0}_{2} \right] +hc
\eea

The neutral gauginos and higgsinos are  mixed  by  matrix $N$
 \be
\tilde \chi^0_{\alpha}  = \sum_{\stackrel {\scriptstyle {\beta
={\rm gauginos},}}{\scriptstyle{{\rm higgsinos}}}} N_{\alpha\beta}
\psi^0_{\beta},\ee where the $\tilde\chi_{\alpha}^0$ forms fifteen
physical Majorana particles. While, the charged gauginos mix with
higgsinos via two mixing matrices $C^\pm_{\alpha\beta}$ \be \tilde
\chi^\pm_{\alpha}  = \sum_{\stackrel {\scriptstyle {\beta ={\rm
gauginos},}}{\scriptstyle{\rm higgsinos}}} C^\pm_{\alpha\beta}
\psi^\pm_{\beta} \ee to form six physical Dirac particles.  These
matrices will be presented in the next section.

\subsection{The Discrete Symmetry for
Proton Stability and Neutrino Masses in SUSY331RN}
 \label{sec:neutrinosmassterms}

As before, if we choose the R-charges as follows\bea
n_{L}&=&n_{\eta}=n_{\chi}= \frac{1}{2}, \crn
n_{\eta^{\prime}}&=&n_{\chi^{\prime}}=n_{u}=n_{u^{\prime}}=-
\frac{1}{2}, \crn n_{Q_{3}}&=&n_{\rho^{\prime}}=1, \ n_{d} =
n_{d^{\prime}}=-2, \crn n_{\rho}&=&-1, \ n_{Q_{\alpha}} =
\frac{3}{2}, \ n_{l}=- \frac{3}{2}, \eea then the terms under this
symmetry are obtained by\bea W&=&\frac{1}{2} \left(
\mu_{0a}\hat{L}_{a} \hat{ \eta}^{\prime}+ \mu_{1a}\hat{L}_{a}
\hat{ \chi}^{\prime}+ \mu_{ \eta} \hat{ \eta} \hat{
\eta}^{\prime}+ \mu_{ \chi} \hat{ \chi} \hat{ \chi}^{\prime}+
\mu_{2} \hat{\eta} \hat{ \chi}^{\prime}+ \mu_{3} \hat{\chi} \hat{
\eta}^{\prime}+ \mu_{ \rho} \hat{ \rho} \hat{ \rho}^{\prime}
\right) \crn & & + \frac{1}{3} \left( \lambda_{1ab} \hat{L}_{a}
\hat{ \rho}^{\prime} \hat{l}^{c}_{b}+ \lambda_{2a} \epsilon
\hat{L}_{a} \hat{\chi} \hat{\rho}+ \lambda_{3a} \epsilon
\hat{L}_{a} \hat{\eta} \hat{\rho}+ \lambda_{4ab} \epsilon
\hat{L}_{a} \hat{L}_{b} \hat{\rho}+ \kappa_{1i} \hat{Q}_{3}
\hat{\eta}^{\prime} \hat{u}^{c}_{i}+ \kappa_{1}^{\prime}
\hat{Q}_{3} \hat{\eta}^{\prime} \hat{u}^{\prime c} \right. \crn &
&+ \left. \kappa_{2i} \hat{Q}_{3} \hat{\chi}^{\prime}
\hat{u}^{c}_{i}+ \kappa_{2}^{\prime} \hat{Q}_{3}
\hat{\chi}^{\prime} \hat{u}^{\prime c}+ \kappa_{3\alpha i}
\hat{Q}_{\alpha} \hat{\eta} \hat{d}^{c}_{i}+ \kappa_{3\alpha
\beta}^{\prime} \hat{Q}_{\alpha} \hat{\eta} \hat{d}^{\prime
c}_{\beta}+ \kappa_{4\alpha i}
\hat{Q}_{\alpha}\hat{\rho}\hat{u}^{c}_{i}+
\kappa_{4\alpha}^{\prime} \hat{Q}_{\alpha}\hat{\rho}
\hat{u}^{\prime c} \right. \crn & &+ \left. \kappa_{5i}\hat{Q}_{3}
\hat{\rho}^{\prime} \hat{d}^{c}_{i}+ \kappa_{5
\beta}^{\prime}\hat{Q}_{3} \hat{\rho}^{\prime}
\hat{d}^{c}_{\beta}+ \kappa_{6\alpha i} \hat{Q}_{\alpha}
\hat{\chi} \hat{d}^{c}_{i}+ \kappa_{6\alpha \beta}^{\prime}
\hat{Q}_{\alpha} \hat{\chi} \hat{d}^{\prime c}_{\beta}+ f_{1}
\epsilon \hat{ \rho} \hat{ \chi} \hat{ \eta}+ f^{\prime}_{1}
\epsilon \hat{ \rho}^{\prime} \hat{ \chi}^{\prime}\hat{
\eta}^{\prime} \right. \crn & &+ \left. \lambda^{\prime}_{\alpha
ai}\hat{Q}_{\alpha}\hat{L}_{a} \hat{d}^{c}_{i}+ \xi_{2 \alpha a
\beta}\hat{Q}_{\alpha} \hat{L}_{a} \hat{d}^{\prime c}_{\beta}
\right). \label{neutrinomasses} \eea In this case, it is easy to
see that the fields $L,\ l,\ Q_{\alpha},\ u,\ u^{\prime},\
\tilde{\eta},\ \tilde{\eta}^{\prime},\ \tilde{\chi},\
\tilde{\chi}^{\prime},\  \rho ,\ \rho^{\prime},\ \tilde{Q}_{3},\
\tilde{d}$ and $\tilde{d}^{\prime}$ have $R$-charge equal to one,
while the others fields have opposite $R$-charge.

The  superpotential in (\ref{neutrinomasses}) provides us the mass
terms for leptons and higgsinos \be -
\frac{\mu_{0a}}{2}L_{a}\tilde{\eta}^{\prime}-
\frac{\mu_{1a}}{2}L_{a}\tilde{\chi}^{\prime}-
\frac{\mu_{2}}{2}\tilde{\eta}\tilde{\chi}^{\prime}-
\frac{\mu_{3}}{2}\tilde{\chi}\tilde{\eta}^{\prime}-
\frac{\lambda_{2a}}{3}(L_{a}\tilde{\chi}\rho +
\tilde{\rho}L_{a}\chi)- \frac{\lambda_{3a}}{3}
(L_{a}\tilde{\eta}\rho + \tilde{\rho}L_{a}\eta)-
\frac{\lambda_{4ab}}{3}L_{a}L_{b}\rho + hc, \ee which in terms of
field components  get a form \bea &-& \frac{\mu_{0a}}{2}(
\nu_{a}\tilde{\eta}^{\prime 0}_{1}+l_{a}\tilde{\eta}^{\prime +}+
\nu^{c}_{a}\tilde{\eta}^{\prime 0}_{2})- \frac{\mu_{1a}}{2}(
\nu_{a}\tilde{\chi}^{\prime 0}_{1}+l_{a}\tilde{\chi}^{\prime +}+
\nu^{c}_{a}\tilde{\chi}^{\prime 0}_{2}) \crn &-&
\frac{\mu_{2}}{2}(\tilde{\eta}^{0}_{1}\tilde{\chi}^{\prime
0}_{1}+\tilde{\eta}^{-}\tilde{\chi}^{\prime +}
+\tilde{\eta}^{0}_{2}\tilde{\chi}^{\prime 0}_{2})-
\frac{\mu_{3}}{2}(\tilde{\chi}^{0}_{1}\tilde{\eta}^{\prime
0}_{1}+\tilde{\chi}^{-}\tilde{\eta}^{\prime +}
+\tilde{\chi}^{0}_{2}\tilde{\eta}^{\prime 0}_{2})-
\frac{\lambda_{2a}}{3}[
(\nu^{c}_{a}\tilde{\chi}^{0}_{1}-\nu_{a}\tilde{\chi}^{0}_{2})
\rho^{0} \crn  &+& ( \tilde{\rho}^{0}\nu^{c}_{a}-
\tilde{\rho}^{+}_{2}l_{a}) \chi^{0}_{1}+(
\tilde{\rho}^{+}_{1}l_{a}- \tilde{\rho}^{0}\nu_{a}) \chi^{0}_{2}]-
\frac{\lambda_{3a}}{3}[
(\nu^{c}_{a}\tilde{\eta}^{0}_{1}-\nu_{a}\tilde{\eta}^{0}_{2})
\rho^{0}+ ( \tilde{\rho}^{0}\nu^{c}_{a}-
\tilde{\rho}^{+}_{2}l_{a}) \eta^{0}_{1} \crn &+& (
\tilde{\rho}^{+}_{1}l_{a}- \tilde{\rho}^{0}\nu_{a}) \eta^{0}_{2}]-
\frac{\lambda_{4ab}}{3}( \nu^{c}_{a}\nu_{b}- \nu^{c}_{b}\nu_{a})
\rho^{0}+h.c \eea It is easy to see that the above Lagrangian
gives mass terms for neutrinos and forbids the proton decay.

\section{Fermion masses}
\label{sec:fermionmasses}

With the above mass terms, we get the mass matrices for the
neutral and charged fermions, respectively. Diagonalizing these
matrices, we get the physical masses for the fermions.

\subsection{Masses of the neutral fermions}

Mass Lagrangian for the neutral fermions are easily obtained by
\bea {\cal M}_{neutral}&=&- \frac{1}{2} \left[
\mu_{0a}(\nu_{aL}\tilde{\eta}^{\prime 0}_1+
\nu^c_{aL}\tilde{\eta}^{\prime 0}_2)+\mu_{1a}(
\nu_{aL}\tilde{\chi}^{\prime 0}_{1}
+\nu^c_{aL}\tilde{\chi}^{\prime 0}_2)+
\mu_{\eta}(\tilde{\eta}^{0}_1\tilde{\eta}^{\prime 0}_1+
\tilde{\eta}^{0}_2\tilde{\eta}^{\prime 0}_2) \right. \crn & &+
\left. \mu_{\chi}(\tilde{\chi}^{0}_1\tilde{\chi}^{\prime 0}_1+
\tilde{\chi}^{0}_2\tilde{\chi}^{\prime 0}_2)+
\mu_{\rho}\tilde{\rho}^{0}\tilde{\rho}^{\prime 0}+
\mu_{2}(\tilde{\eta}^{0}_{1}\tilde{\chi}^{\prime
0}_{1}+\tilde{\eta}^{0}_{2}\tilde{\chi}^{\prime 0}_{2})+
\mu_{3}(\tilde{\chi}^{0}_{1}\tilde{\eta}^{\prime
0}_{1}+\tilde{\chi}^{0}_{2} \tilde{\eta}^{\prime 0}_{2}) \right]
\crn & &- \frac{1}{3} \left[
\lambda_{2a}u(\nu^c_{aL}\tilde{\chi}^{0}_1-
\nu_{aL}\tilde{\chi}^{0}_2)
-\lambda_{2a}w\nu_{aL}\tilde{\rho}^{0}+
\lambda_{3a}u(\nu^c_{aL}\tilde{\eta}^{0}_1-
\nu_{aL}\tilde{\eta}^{0}_2) \right. \crn & &+\left. \lambda_{3a}v
\nu^c_{aL}\tilde{\rho}^{0}+ \lambda_{4ab}u(\nu^c_{aL}\nu_{bL}-
\nu^{c}_{bL}\nu_{aL}) +f_1v \tilde{\rho}^{0}\tilde{\chi}^{0}_2
+f_1 w \tilde{\rho}^{0}\tilde{\eta}^{0}_1 \right. \crn & &+ \left.
f_1 u (\tilde{\chi}^{0}_2\tilde{\eta}^{0}_1-\tilde{\chi}^{0}_1
\tilde{\eta}^{0}_2) +f'_1v'\tilde{\rho}^{\prime
0}\tilde{\chi}^{\prime 0}_2+ f'_1 w'\widetilde{\rho}^{\prime
0}\tilde{\eta}^{\prime 0}_1 + f'_1 u'(\tilde{\chi}^{\prime 0}_2
\tilde{\eta}^{\prime 0}_1- \tilde{\chi}^{\prime
0}_1\tilde{\eta}^{\prime 0}_2) \right] \crn &
&-\frac{m_{\lambda}}{2}(2 \lambda_{X^{0}}\lambda_{X^{0*}}+
\lambda^{3}_{A}\lambda^{3}_{A}+ \lambda^{8}_{A}\lambda^{8}_{A})-
\frac{m^{\prime}}{2} \lambda_{B}\lambda_{B} \crn & &
+\frac{igv}{\sqrt{2}}\tilde{\eta}^{0}_{1} \lambda^{3}_{A}+igv
\tilde{\eta}^{0}_{2}\lambda_{X^{0}}+
\frac{igv}{\sqrt{6}}\tilde{\eta}^{0}_{1} \lambda^{8}_{A} +igw
\tilde{\chi}^{0}_{1}\lambda_{X^{0*}}\crn & &- igw
\sqrt{\frac{2}{3}}\tilde{\chi}^{0}_{2} \lambda^{8}_{A}-
\frac{igu}{\sqrt{2}}\tilde{\rho}^{0} \lambda^{3}_{A}+
\frac{igu}{\sqrt{6}}\tilde{\rho}^{0} \lambda^{8}_{A}-
\frac{igv^{\prime}}{\sqrt{2}}\tilde{\eta}^{\prime 0}_{1}
\lambda^{3}_{A}-igv^{\prime} \tilde{\eta}^{\prime
0}_{2}\lambda_{X^{0*}} \crn & &-
\frac{igv^{\prime}}{\sqrt{6}}\tilde{\eta}^{\prime 0}_{1}
\lambda^{8}_{A}-igw^{\prime} \tilde{\chi}^{\prime
0}_{1}\lambda_{X^{0}}+igw^{\prime}
\sqrt{\frac{2}{3}}\tilde{\chi}^{\prime 0}_{2} \lambda^{8}_{A}-
\frac{igu^{\prime}}{\sqrt{6}}\tilde{\rho}^{\prime 0}
\lambda^{8}_{A}- \frac{igu^{\prime}}{\sqrt{2}}\tilde{\rho}^{\prime
0} \lambda^{3}_{A} \crn & &+ig^{\prime}u \sqrt{\frac{2}{3}}
\tilde{\rho}^{0}\lambda_{B}- \frac{ig^{\prime}v}{3
\sqrt{2}}\tilde{\eta}^{0}_{1} \lambda_{B} - \frac{ig^{\prime}w}{3
\sqrt{2}}\tilde{\chi}^{0}_{2}\lambda_{B}+\frac{ig^{\prime}v^{\prime}}{3
\sqrt{2}}\tilde{\eta}^{\prime 0}_{1}\lambda_{B} +
\frac{ig^{\prime}w^{\prime}}{3 \sqrt{2}}\tilde{\chi}^{\prime
0}_{2}\lambda_{B} \crn & &-ig^{\prime}u^{\prime}
\sqrt{\frac{2}{3}} \tilde{\rho}^{0}\lambda_{B}+h.c. \nn \eea
 In the basis $\Psi^0$ of the form
\bea \left(\nu_{1}\mbox{ }\nu_{2}\mbox{ } \nu_{3}\mbox{
}\nu^{c}_{1}\mbox{ } \nu^{c}_{2}\mbox{ } \nu^{c}_{3}-i
\lambda^{3}_{A}-i \lambda_{X^{0}}-i \lambda_{X^{0*}}-i
\lambda^{8}_{A}-i \lambda_{B}\mbox{ } \tilde{\eta}^0_1\mbox{
}\tilde{\eta}^{\prime 0}_1\mbox{ } \tilde{\eta}^0_2\mbox{ }
\tilde{\eta}^{\prime 0}_2\mbox{ }\tilde{\chi}^0_1\mbox{ }
\tilde{\chi}^{\prime 0}_1\mbox{ } \tilde{\chi}^0_2\mbox{
}\tilde{\chi}^{\prime 0}_2\mbox{ }\tilde{\rho}^0\mbox{ }
\tilde{\rho}^{\prime 0}\right),\nn \eea  the mass Lagrangian can
be written as follows \be - \frac{1}{2}(
\Psi^{0})^{T}Y^{0}\Psi^{0}+h.c.\ee Here $Y^0$ is symmetric matrix
with the nonzero elements given in Appendix.\ref{neumatel}, where
the mass eigenstates are given by \bea \tilde{
\chi}^{0}_{i}&=&N_{ij} \Psi^{0}_{j}, \,\ j=1, \cdots ,21.
\label{emasneu}\eea

The mass matrix of the neutral fermions consists of three parts:
(a) The first part $M_{\nu}$ is the $6\times 6$ mass matrix of the
neutrinos which belongs to the 3-3-1 model with right-handed
neutrinos; (b) While, the second part $M_{N}$ is the $15\times 15$
mass matrix of the neutralinos which exists only in its
supersymmetric version; (c) The last part $M_{\nu N}$ arises due
to mixing among the neutrinos and neutralinos. Thus, the mass matrix
for the neutral fermions is signified as follows\be Y^0=\left(%
\begin{array}{cc}
  M_\nu & M_{\nu N} \\
  M^T_{\nu N} & M_N \\
\end{array}%
\right).\ee

To keep consistent with the low-energy effective theories, some
Yukawa couplings in the sub-matrices will be fixed in terms of
fine-tunings needed in this model. First, we know that the mass
matrix $M_\nu$ gives three Dirac eigenstates. Two of them have
degenerate eigenvalues
$m_\nu=\fr{2u}{3}\sqrt{\la^2_{412}+\la^2_{413}+\la^2_{423}}$ and
the other one massless. Thus, it is easy to identify the mass
splitting $m_\nu$ as the value of measured atmospheric neutrino
mass difference $\Delta m_{atm}\sim 5\times 10^{-2}$ ~eV. By
putting $\la_{412}=\la_{413}=\la_{423}$, we get
\begin{eqnarray}
\la_{412}&=&\la_{413} =\la_{423}= 2\times 10^{-13},\crn
\la_{421}&=&\la_{431}= \la_{432}= -2\times 10^{-13}.
\end{eqnarray}
It is worth emphasizing that when the mixing terms turn on, this
inverted spectrum will not only give rise to mass splitting
between the two degenerate Dirac states, it will also split each
Dirac pairs into two non-degenerate Majorana states, resulting in
the spectrum with six Majorana eigenstates with four heavier ones
and two light ones. Here we are assuming that the solar
oscillation is between the two heavier Majorana states.

Finally, to keep the mass constraints from astrophysics and
cosmology \cite{pdg} as well as to be consistent with all the
earlier analyses \cite{lepmass}, the parameters in the mass matrix
$M_N$ can be chosen as a typical example by:  \bea
\mu_{\eta}&=&300\ \mathrm{GeV},\ \mu_{\rho}=500\ \mathrm{GeV},\
\mu_{\chi}=700\ \mathrm{GeV},\crn \mu_2&=&50\ \mathrm{GeV},\
\mu_3=200\ \mathrm{GeV},\ m^\prime=2000\ \mathrm{GeV},\crn
\mu_{\la}&=&3000\ \mathrm{GeV},\ f_1=1.8,\
f^{\prime}_1=10^{-3}.\label{efes}\label{para} \eea Here in this
model, the Higgs bosons' VEVs are fixed as follows\bea
v_{\eta}&=&15\ \mathrm{GeV},\ v_{\eta'}=10\ \mathrm{GeV}, \crn
v_{\rho}&=&244.9\ \mathrm{GeV},\ v_{\rho'}=13\ \mathrm{GeV}, \crn
v_{\chi_2}&=&v_{\chi'_2}=1000\ \mathrm{GeV}, \label{vevs} \eea and
the value of $g$ given in \cite{pdg}.

Now, the mixing terms rise to correct the effective ones, thus, in
some ways we can get the physical masses. In the first case, the
parameters in the mixing matrix $M_{\nu N}$ can be chosen as
follows\ben \item For the dimensionless parameters:
\bea\la_{21}&=&0,\ \la_{22}=0,\ \la_{23}=0, \crn \la_{31}&=&0,\
\la_{32}=0,\ \la_{33}=0, \label{lambdas} \eea \item For the
mass-scale parameters (in GeV): \bea \mu_{01}&=&10^{-4},\
\mu_{02}=10^{-4},\ \mu_{03}=0,\crn \mu_{11}&=&10^{-5},\
\mu_{12}=0,\ \mu_{13}=0.\label{mus0} \eea\een Hence, the
neutralinos obtain the masses in GeV as follows
\begin{eqnarray}
&& 3210.2199,\ 3152.6352,\ 3152.6352,\ 3004.7008,\ 2087.1869,\crn
&& 693.1140,\ 672.9818,\ 367.2895,\ 281.0245,\ 269.3481,\crn &&
151.7631,\ 115.9158,\ 48.4871,\ 46.7180,\ 39.5593.\label{hnm}
\end{eqnarray}
The tauon neutrino gains masses in eV by
\begin{equation}
5.39600\times 10^{-5},\ 4.12596\times 10^{-7},\label{mnus1}
\end{equation}
while the masses of the electron and muon neutrinos are (in eV)
\begin{eqnarray}
&&6.31927\times 10^{-2}, \ 5.49914\times 10^{-2}, \crn
&&5.46530\times 10^{-2}, \ 4.70848\times 10^{-2}.\label{mnus2}
\end{eqnarray}

In the second case, the values of the parameters in the mixing
matrix $M_{\nu N}$ are given in another way as below \ben \item
For the mass-scale parameters: \bea \mu_{01}&=&0,\ \mu_{02}=0,\
\mu_{03}=0,\crn \mu_{11}&=&0,\ \mu_{12}=0,\ \mu_{13}=0,\eea \item
For the dimensionless parameters: \bea \la_{21}&=&10^{-6},\
\la_{22}=10^{-6},\ \la_{23}=10^{-6},\crn \la_{31}&=&10^{-6},\
\la_{32}=0,\ \la_{33}=0. \eea\een The masses (in GeV) of the
neutralinos are \bea && 3210.2199,\ 3152.6352,\ 3152.6352,\
3004.7008,\ 2087.1869,\crn && 693.1140,\ 672.9818,\ 367.2895,\
281.0245,\ 269.3481,\crn && 151.7631,\ 115.9158,\ 48.4871,\
46.7180,\ 39.5593.\eea The tauon neutrino in this case gains
masses in eV by \be 1.95303\times 10^{-4},\ 4.87078\times
10^{-5}.\ee Also, the electron and muon neutrinos have masses in
eV \bea &&8.37329\times 10^{-2},\ 5.52047\times 10^{-2},\crn
&&5.51605\times 10^{-2},\ 3.69629\times 10^{-2}.\eea Notice that
the coupling constant $g^{\prime},\lambda_{4ab}$ and the parameter
$m^\prime$ appear only in the mass matrix of the neutral fermions,
while, those of the charged sector are $\la_{1ab}$.

\begin{figure}
\begin{center}
\begin{picture}(100,150)(0,0)
\Line(10,30)(90,30) \Line(10,40)(90,40) \Line(10,120)(90,120)
\Line(10,130)(90,130)
 \Line(10,160)(90,160)
\Line(10,170)(90,170) \Line(45,35)(45,125)
 \Line(55,125)(55,165)
\Text(90,80)[]{atmospheric } \Text(72,145)[]{solar }
 \Text(0,30)[]{$m_5$}
 \Text(0,40)[]{$m_6$}
\Text(0,120)[]{$m_3$}
 \Text(0,130)[]{$m_4$}
 \Text(0,160)[]{$m_1$}
 \Text(0,170)[]{$m_2$}
\end{picture}
\caption[]{ The inverted hierarchy mass pattern of  neutrinos in
the model.} \label{fig}
\end{center}
\end{figure}
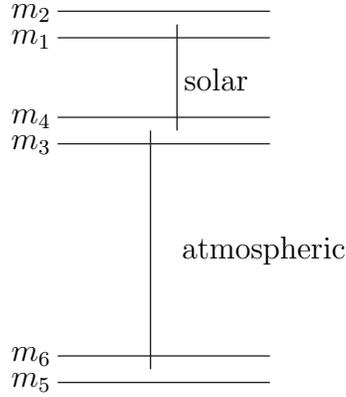

As above, we give two typical examples of the values of the mixing
terms which not only give the consistent mass spectra of the
neutrinos but also keep an large enough hierarchy so that the
neutralinos gain the masses satisfying the lower mass limit
($>32.5$ GeV) from astrophysics and cosmology. Consequently, the
neutrinos in this model yield inverted hierarchy mass patterns as
shown in Fig.(\ref{fig}).

\subsection{Masses of the charged fermions}

The terms  contributing to the masses of the charged fermions are
\bea {\cal M}_{charged}&=&- \frac{\lambda_{1ab}}{3}
l_{a}l^{c}_{b}u'-\frac{f_{1}}{3}(\tilde{\rho}^{+}_{2}\tilde{\chi}^{-}v
+ \tilde{\rho}^{+}_{1}\tilde{\eta}^{-}w)-
\frac{f^{\prime}_{1}}{3}(
 \tilde{\rho}^{\prime -}_{2}\tilde{\chi}^{\prime +}v^{\prime}+
\tilde{\rho}^{\prime -}_{1}\tilde{\eta}^{\prime +}w^{\prime}) \crn
& &- m_{\lambda}( \lambda^{-}_{W}\lambda^{+}_{W}+
\lambda^{-}_{Y}\lambda^{+}_{Y}) +ig[v
\tilde{\eta}^{-}\lambda^{+}_{W}+w
\tilde{\chi}^{-}\lambda^{+}_{Y}+u(
\tilde{\rho}^{+}_{1}\lambda^{-}_{W}+\tilde{\rho}^{+}_{2}\lambda^{-}_{Y})-
v^{\prime} \tilde{\eta}^{\prime +}\lambda^{-}_{W} \crn &
&-w^{\prime} \tilde{\chi}^{\prime +}\lambda^{-}_{Y}- u^{\prime}(
\tilde{\rho}^{\prime -}_{1}\lambda^{+}_{W}+\tilde{\rho}^{\prime
-}_{2}\lambda^{+}_{Y})]- \frac{\mu_{\eta}}{2} \tilde{\eta}^{-}
\tilde{\eta}^{\prime +}
-\frac{\mu_{\rho}}{2}(\tilde{\rho}^{+}_{1}\tilde{\rho}^{\prime
-}_{1}+ \tilde{\rho}^{+}_{2}\tilde{\rho}^{\prime -}_{2}) -
\frac{\mu_{\chi}}{2}\tilde{\chi}^{-}\tilde{\chi}^{\prime +} \crn &
&- \fr{\mu_{0a}}{2}l_a\tilde{\eta}'^+-\fr{\mu_{1a}}
{2}l_a\tilde{\chi}'^+-
\frac{\mu_{2}}{2}\tilde{\eta}^{-}\tilde{\chi}^{\prime +}-
\frac{\mu_{3}}{2}\tilde{\chi}^{-}\tilde{\eta}^{\prime +}
-\frac{\lambda_{2a}}{3}l_{a}\tilde{\rho}^{+}_{1} w
+\fr{\la_{3a}}{3} l_{a}\tilde{\rho}^{+}_{2} v +h.c. \nn \eea
 To write the mass matrix, we will choose the following bases \bea
\psi^{-}&=&\left( \begin{array}{ccccccccc} l_{1} & l_{2} & l_{3}
&-i \lambda^{-}_{W}&-i \lambda^{-}_{Y}& \tilde{\eta}^{-}&
\tilde{\chi}^{-}& \tilde{\rho}^{\prime -}_{1}&
\tilde{\rho}^{\prime -}_{2}
\end{array}
\right)^{T} \crn \psi^{+}&=&\left( \begin{array}{ccccccccc}
l_{1}^{c} & l_{2}^{c} & l_{3}^{c} &i \lambda^{+}_{W}&i
\lambda^{+}_{Y}& \tilde{\eta}^{\prime +}& \tilde{\chi}^{\prime +}&
\tilde{\rho}^{+}_{1}& \tilde{\rho}^{+}_{2}
\end{array}
\right)^{T} \eea and define
\be
\Psi^{\pm}=\left(\begin{array}{cc}
\psi^{+} & \psi^{-}
\end{array}\right)^{T}.
\ee With these definitions, the mass term is written in the form
\be - \frac{1}{2}( \Psi^{\pm})^{T}Y^{\pm}\Psi^{\pm}+h.c., \ee
where \be Y^{\pm}= \left( \begin{array}{cc}
0 & X^{T} \\
X & 0
\end{array}\right).
\ee Then, the $X$ matrix is given by
\bea X= \left(%
\begin{array}{ccccccccccc}
  \fr{\la_{111}}{3}u' & \fr{\la_{112}}{3}u' &
  \fr{\la_{113}}{3}u' & 0 & 0 & \fr{1}{2}\mu_{01} &
   \fr{1}{2}\mu_{11} & \fr{\la_{21}}{3}w & -\fr{\la_{31}}{3}v  \\
  \fr{\la_{121}}{3}u' & \fr{\la_{122}}{3}u' &
  \fr{\la_{123}}{3}u' & 0 & 0 & \fr{1}{2}\mu_{02} &
  \fr{1}{2}\mu_{12} & \fr{\la_{22}}{3}w & -\fr{\la_{32}}{3}v  \\
  \fr{\la_{131}}{3}u' & \fr{\la_{132}}{3}u' &
  \fr{\la_{133}}{3}u' & 0 & 0 & \fr{1}{2}\mu_{03} &
  \fr{1}{2}\mu_{13} & \fr{\la_{23}}{3}w & -\fr{\la_{33}}{3}v  \\
  0 & 0 & 0 &- m_{\la} & 0 &- gv' & 0 & gu & 0  \\
  0 & 0 & 0 & 0 &- m_{\la}  & 0 &- gw' & 0 & gu \\
  0 & 0 & 0 & gv & 0 & \fr{\mu_{\eta}}{2} &
  \frac{\mu_{2}}{2} & \fr{f_1}{3}w & 0 \\
  0 & 0 & 0 & 0 & gw & \frac{\mu_{3}}{2} &
  \fr{\mu_{\chi}}{2} & 0 & \fr{f_1}{3}v \\
  0 & 0 & 0 &- gu' & 0 & \fr{f'_1}{3}w' &
   0 & \fr{\mu_{\rho}}{2} & 0 \\
  0 & 0 & 0 & 0 &- gu' & 0 & \fr{f'_1}{3}v' &
  0 & \fr{\mu_{\rho}}{2} \\
\end{array}%
\right).\nn\eea
 The chargino mass matrix $Y^\pm$ is diagonalized by
using two unitary matrices, $D$ and $E$, defined by \bea \tilde{
\chi}^{+}_{i}=D_{ij} \Psi^{+}_{j}, \,\ \tilde{
\chi}^{-}_{i}=E_{ij} \Psi^{-}_{j}, \,\ i,j=1, \cdots , 9.
\label{2sc} \eea
 The characteristic equation for the  matrix
$Y^{\pm}$  is \bea \det (Y^{\pm}- \lambda I)= \det \left[ \left(
\begin{array}{cc}
- \lambda & X^{T} \\
X  &- \lambda
\end{array} \right) \right]= \det( \lambda^{2}-X^{T} \cdot X).
\label{propmat1} \eea
  Since $X^{T} \cdot X$ is a symmetric
matrix, $\lambda^2$ must be real and positive because $Y^{\pm}$ is
also symmetric. Hence, to obtain eigenvalues,  one only have to
calculate $X^{T} \cdot X$. Then we can write the diagonal mass
matrix as \be M_{SCM}=E^{*}XD^{-1}. \label{m1} \ee
 To determine $E$ and $D$, it is useful the following observation
\be M^{2}_{SCM}=DX^T \cdot XD^{-1}=E^{*}X \cdot X^T(E^{*})^{-1},
\label{m2} \ee which means that $D$ diagonalizes $X^{T} \cdot X$,
while $E^{*}$ diagonalizes $X \cdot X^{T}$. In this case we can
define the following Dirac spinors: \bea \Psi(\tilde{
\chi}^{+}_{i})= \left(
\begin{array}{cc}
             \tilde{ \chi}^{+}_{i} &
         \bar{ \tilde{\chi}}^{-}_{i}
\end{array} \right)^T, \,\
\Psi^{c}(\tilde{ \chi}^{-}_{i})= \left( \begin{array}{cc}
             \tilde{ \chi}^{-}_{i} &
         \bar{ \tilde{\chi}}^{+}_{i}
\end{array} \right)^T.
\label{emasssim} \eea where $\tilde{ \chi}^{+}_{i}$ is the
particle and $\tilde{ \chi}^{-}_{i}$ is the
anti-particle~\cite{mssm,mcr}.

Now to get mass values, all the parameters in the neutral sector
should be kept in this sector of the charged fermions.
Corresponding to the first case in the neutral sector, we have
obtained the following masses (in GeV) for the charginos: \be
3156.1474,\ 3004.4371,\ 665.1171,\ 263.3014,\ 209.0726,
45.9104.\label{cm} \ee The ordinary leptons gain masses in GeV as
$m_\tau=1.7766$, $m_\mu=0.1057$, $m_e=0.00051$. In this case, the
masses have been obtained by using the remaining set of the
dimensionless parameters: \bea \la_{111}&=&1.18\times 10^{-4},\hs
\la_{112}=10^{-7},\hs \la_{113}=10^{-7},\crn
\la_{121}&=&10^{-7},\hs \la_{122}=2.44\times 10^{-2},\hs
\la_{123}=10^{-7},\crn \la_{131}&=&10^{-7},\hs
\la_{132}=10^{-7},\hs \la_{133}=4.10\times 10^{-1}.\eea The second
case is obtained by changing only the mixing terms as in the
neutral sector. As a result, the masses are {\it the same} as in
the previous case. Thus, the ordinary charged leptons get the
consistent masses; and, the lightest chargino with the mass of
$45.9104$ GeV is in the experimental lower limit of $45$ GeV. It
was shown in this section that, there are nine fermions. However,
as mentioned above,  by the conservation of R-parity, there are
only six charginos.

To summarize, as above we have given at the tree level the
consistent masses for the charged leptons and the neutrinos in the
supersymmetric 3-3-1 model with right-handed neutrinos. Such a
model for the leptons is more simpler than that of the
supersymmetric minimal 3-3-1 model \cite{331susy}, where the loop
corrections are needed for the masses of the leptons
\cite{lepmass}. The charginos and neutralinos in this model gain
the masses respectively very smaller than those of the
supersymmetric minimal 3-3-1 model \cite{lepmass}. Contrasting
with the supersymmetric minimal 3-3-1 model, the MSSM neutralinos
and charginos in this model can be directly identified via the
mass spectra given above, where the mass constraints on the MSSM
particles can be found in Ref.\cite{dat}.

\section{Conclusion}

In this article, we have found, in framework of the supersymmetric
3-3-1 model with right-handed neutrinos, definitions of the $R$
charges which are similar with that in the MSSM. This means that
in the considered  model, there is  one discrete symmetry which
allows the neutral and charged fermions, gauge bosons and scalar
fields to get masses and at same time forbidding the proton decay.
Thus, in this case there exists one phenomenology similar to  the
MSSM  with the famous missing transverse energy events, which is
specific of the  $R$-parity conservation \cite{dress}.

We have showed that  there is one symmetry which  gives neutrinos
masses  but forbids the proton decay.  Unlike the cases with the
MSSM and the minimal 3-3-1 model~\cite{lepmass}, in this model,
all the fermions get masses at the tree level.

The famous relation for the R-parity in the MSSM  has been
generalized to this kind of the 3-3-1 models. In this case it
relates to the new conserved charge $\mathcal{L}$. A simple
mechanism for the mass generation of the neutrinos has been
explored. We have showed that the model naturally gives rise to
the neutrinos an inverted hierarchy mass pattern. Moreover, the
MSSM superpartners in this model can be explicitly identified
which is unlike in the case of  supersymmetric extension of the
minimal 3-3-1 model.

\section*{Acknowledgments}

M. C. R would like to thank Vietnam Academy of Science and
Technology by the nice hospitality, warm atmosphere during his
stay at Institute of Physics to start this work.

\appendix

\section{Mass matrix  elements of $Y^{0}$}
\label{neumatel}

The nonzero elements of $Y^{0}$ are \bea m_{1,5}&=&m_{5,1}=-
\frac{u}{3}( \lambda_{421}- \lambda_{412}), \,\ m_{1,6}=m_{6,1}=-
\frac{u}{3}( \lambda_{431}- \lambda_{413}), \crn
m_{1,13}&=&m_{13,1}= \frac{\mu_{01}}{2}, m_{1,14}=m_{14,1}=-
\frac{\lambda_{31}}{3}u, \,\ m_{1,17}=m_{17,1}=
\frac{\mu_{11}}{2}, \crn m_{1,18}&=&m_{18,1}=-
\frac{\lambda_{21}}{3}u, \,\ m_{1,20}=m_{20,1}=-
\frac{\lambda_{21}}{3}w, \crn m_{2,4}&=&m_{4,2}=- \frac{u}{3}(
\lambda_{412}- \lambda_{421}), \,\ m_{2,6}=m_{6,2}=- \frac{u}{3}(
\lambda_{432}- \lambda_{423}), \crn m_{2,13}&=&m_{13,2}=
\frac{\mu_{02}}{2}, \,\ m_{2,14}=m_{14,2}=-
\frac{\lambda_{32}}{3}u, \,\ m_{2,17}=m_{17,2}=
\frac{\mu_{12}}{2}, \crn  m_{2,18}&=&m_{18,2}=-
\frac{\lambda_{22}}{3}u, \,\ m_{2,20}=m_{20,2}=-
\frac{\lambda_{22}}{3}w, \crn  m_{3,4}&=&m_{4,3}=- \frac{u}{3}(
\lambda_{413}- \lambda_{431}), \,\ m_{3,5}=m_{5,3}=- \frac{u}{3}(
\lambda_{423}- \lambda_{432}), \crn m_{3,13}&=&m_{13,3}=
\frac{\mu_{03}}{2}, \,\ m_{3,14}=m_{14,3}=-
\frac{\lambda_{33}}{3}u, \,\ m_{3,17}=m_{17,3}=
\frac{\mu_{13}}{2}, \crn  m_{3,18}&=&m_{18,3}=-
\frac{\lambda_{23}}{3}u, \,\ m_{3,20}=m_{20,3}=-
\frac{\lambda_{23}}{3}w, \crn  m_{4,12}&=&m_{12,4}=
\frac{\lambda_{31}}{3}u, \,\ m_{4,15}=m_{15,4}=
\frac{\mu_{01}}{2}, \,\ m_{4,16}=m_{16,4}=
\frac{\lambda_{21}}{3}u, \crn  m_{4,19}&=&m_{19,4}=
\frac{\mu_{11}}{2}, \,\ m_{4,20}=m_{20,4}=
\frac{\lambda_{31}}{3}v, \crn  m_{5,12}&=&m_{12,5}=
\frac{\lambda_{32}}{3}u, \,\ m_{5,15} =m_{15,5}=
\frac{\mu_{02}}{2}, \,\ m_{5,16}=m_{16,5}=
\frac{\lambda_{22}}{3}u, \crn  m_{5,19}&=&m_{19,5}=
\frac{\mu_{12}}{2}, \,\ m_{5,20}=m_{20,5}=
\frac{\lambda_{32}}{3}v, \crn  m_{6,12}&=&m_{12,6}=
\frac{\lambda_{33}}{3}u, \,\ m_{6,15}= m_{15,6}=
\frac{\mu_{03}}{2}, \,\ m_{6,16}=m_{16,6}=
\frac{\lambda_{23}}{3}u, \crn  m_{6,19}&=&m_{19,6}=
\frac{\mu_{13}}{2}, \,\ m_{6,20}=m_{20,6}=
\frac{\lambda_{33}}{3}v, \crn  m_{7,7}&=&-m_{\lambda}, \,\
m_{7,12}=m_{12,7}= \frac{gv}{\sqrt{2}}, \,\ m_{7,13}=m_{13,7}=-
\frac{gv^{\prime}}{\sqrt{2}}, \crn  m_{7,20}&=&m_{20,7}=-
\frac{gu}{\sqrt{2}}, \,\ m_{7,21}=m_{13,7}=-
\frac{gu^{\prime}}{\sqrt{2}}, \crn
m_{8,9}&=&m_{9,8}=m_{10,10}=-m_{\lambda}, \,\ m_{8,14}=
m_{14,8}=gv, \,\ m_{8,17}=m_{17,8}=-gw^{\prime}, \crn
m_{9,15}&=&m_{15,9}=-gv^{\prime}, \,\ m_{9,16}=m_{16,9}=gw, \crn
m_{10,12}&=&m_{12,10}= \frac{gv}{\sqrt{6}}, \,\
m_{10,13}=m_{13,10}=- \frac{gv^{\prime}}{\sqrt{6}}, \crn
m_{10,18}&=&m_{18,10}=- \sqrt{\frac{2}{3}}gw, \,\
m_{10,19}=m_{19,10}= \sqrt{\frac{2}{3}}gw^{\prime}, \crn
m_{10,20}&=&m_{20,10}= \frac{gu}{\sqrt{6}}, \,\
m_{10,21}=m_{21,10}=- \frac{gu^{\prime}}{\sqrt{6}}, \crn
m_{11,11}&=&-m^{\prime}, \,\ m_{11,12}=m_{12,11}= -
\frac{g^{\prime}v}{3 \sqrt{2}}, \,\ m_{11,13}=m_{13,11}=
\frac{g^{\prime} v^{\prime}}{3 \sqrt{2}}, \crn
m_{11,18}&=&m_{18,11}=- \frac{g^{\prime}w}{3 \sqrt{2}}, \,\
m_{11,19}=m_{19,11}= \frac{g^{\prime}w^{ \prime}}{3 \sqrt{2}},
\crn m_{11,20}&=&m_{20,11}= \sqrt{\frac{2}{3}} g^{\prime}u, \,\
m_{11,21}=m_{21,11}=- \sqrt{\frac{2}{3}} g^{\prime}u^{\prime},
\crn  m_{12,13}&=&m_{13,12}= \frac{\mu_{\eta}}{2}, \,\
m_{12,17}=m_{17,12}= \frac{\mu_{2}}{2}, \crn
m_{12,18}&=&m_{18,12}= \frac{f_{1}}{3}u, \,\ m_{12,20}=m_{20,12}=
\frac{f_{1}}{3}w, \crn  m_{13,16}&=&m_{16,13}= \frac{\mu_{3}}{2},
m_{13,19}=m_{19,13}= \frac{f^{\prime}_{1}}{3}u^{\prime}, \,\
m_{13,21}=m_{21,13}= \frac{f^{\prime}_{1}}{3}w^{ \prime}, \crn
m_{14,15}&=&m_{15,14}= \frac{\mu_{\eta}}{2}, \,\
m_{14,16}=m_{16,14}=- \frac{f_{1}}{3}u, \,\ m_{14,19}=m_{19,14}=
\frac{\mu_{2}}{2}, \crn  m_{15,17}&=&m_{17,15}=-
\frac{f^{\prime}_{1}}{3}u^{\prime}, \,\ m_{15,18}=m_{18,15}=
\frac{\mu_{3}}{2}, \crn m_{16,17}&=&m_{17,16}=
\frac{\mu_{\chi}}{2}, \crn  m_{17,19}&=&m_{19,17}=
\frac{\mu_{\chi}}{2}, \,\ m_{17,20}=m_{20,17}= \frac{f_{1}}{3}u,
\crn  m_{18,19}&=&m_{19,18}= \frac{\mu_{\chi}}{2}, \,\
m_{18,21}=m_{21,18}= \frac{f_{1}}{3}v, \crn m_{19,21}&=&m_{21,19}=
\frac{f^{\prime}_{1}}{3}u^{ \prime}, \crn  m_{20,21}&=&m_{21,20}=
\frac{\mu_{\rho}}{2}. \label{mmn} \eea

\end{document}